%% file: MainText.tex
\documentclass[10pt, article, twocolumn]{article} 

\usepackage[T1]{fontenc}
\usepackage{geometry}
\usepackage{setspace}
\usepackage{physics}
\usepackage{graphicx}
\usepackage{float}
\usepackage[utf8]{inputenc}
\usepackage[style = phys, articletitle=true]{biblatex}
\DeclareUnicodeCharacter{2212}{\textendash}

\usepackage[colorlinks=true, citecolor=blue]{hyperref}
\usepackage{lettrine}
\usepackage{subcaption}
\usepackage{authblk}
\usepackage{subfiles}
\DeclareCaptionFormat{custom}
{%
    \textbf{#1#2}\textit{\small #3}
}
\hypersetup{ colorlinks=true, linkcolor=red, filecolor=black, urlcolor=black }

\title{Wavelength-Dependent Electrical Readout of Spin Ensembles in a Thin-Film SiC-on-Insulator Platform} 

\author[1]{Alexander Zappacosta*}
\author[1]{Ben Haylock}
\author[1]{Paul Fisher}
\author[2, 3]{Naoya Morioka}
\author[1]{Robert Cernansky*}
\affil[1]{ Institute for Quantum Optics, Ulm University, Ulm, Germany}
\affil[2]{Institute for Chemical Research, Kyoto University, Uji, Japan}
\affil[3]{Center for Spintronics Research Network, ICR, Kyoto University, Uji, Japan \newline}
\affil[ ]{*Email: alexander.zappacosta@uni-ulm.de, robert.cernansky@uni-ulm.de}
\affil[ ]{ }
\affil[ ]{Link to publication: \url{https://doi.org/10.1021/acs.nanolett.5c05971}}
\date{\today}

\begin{document}

\twocolumn[ 
  \begin{@twocolumnfalse} 
\maketitle
\begin{abstract}
We report electrical spin state readout and coherent control of an ensemble ($\sim$540) of silicon vacancies ($\mathrm{V}_{\mathrm{Si}}^{-}$) in a silicon carbide-on-insulator (SiCOI) platform, with excitation wavelengths from 780 to 990 nm, demonstrating for the first time spin state readout well beyond the zero phonon line of the V2 $\mathrm{V}_{\mathrm{Si}}^{-}$. By implementing photoelectrical detection of magnetic resonance in thin-film SiCOI, we merge a scalable spin readout technique requiring no collection optics, together with a promising platform for future scalable and CMOS-compatible integrated photonics. Furthermore, we provide a comparison of optical and electrical readout between bulk silicon carbide (SiC) and thin-film SiCOI, revealing that our thin-film processing has a measured $T_2$ coherence time of $\approx$ 7 µs, similar to that in the bulk SiC. These results extend the capabilities of SiCOI toward electronic and spin-based devices for scalable quantum technologies over a wide range of excitation wavelengths.\\

\end{abstract} 
\noindent{\bfseries Keywords}\\

Silicon carbide, quantum electronics, photoelectric detection, silicon vacancy, wavelength dependence, thin-film silicon carbide on insulator \\
  \end{@twocolumnfalse}  
] 

\lettrine[findent=2pt]{\textbf{S}}{ }ilicon 
carbide (SiC) is a CMOS compatible semiconducting material that hosts a variety of solid-state spin systems including the divacancy \cite{wolfowicz_optical_2017, christle_isolated_2015, baranov_epr_2005}, silicon vacancy ($\mathrm{V}_\mathrm{Si}^{-}$) \cite{ widmann_coherent_2015, radulaski_scalable_2017, mizuochi_continuous-wave_2002} and nitrogen vacancy \cite{wang_experimental_2020, norman_sub-2_2025, wang_coherent_2020}. These systems allow for optical initialisation, control, and readout with coherence times exceeding 1 ms \cite{christle_isolated_2015}, making them excellent candidates for a variety of quantum technologies \cite{nagy_high-fidelity_2019, castelletto_silicon_2020} including sensing \cite{fisher_high-resolution_2025,li_non-invasive_2025, simin_all-optical_2016, tahara_quantum_2025, castelletto_quantum_2024, kraus_magnetic_2014} and communication \cite{hu_room-temperature_2024, xu_silicon_2021}. Current bulk SiC experiments utilise optically detected magnetic resonance (ODMR), limiting the scalability due to large amounts of photon collection optics including wavelength filtering and photodiodes.\\
\indent Photoelectrical detection of magnetic resonance (PDMR) \cite{bourgeois_photoelectric_2015} enables full integration of detection directly onto the SiC surface, extracting charges from ionised spin systems for electrical spin state readout. This is possible with patterned electrodes, because their physical size can bypass the optical diffraction limit using standard lithography techniques, making PDMR an excellent solution for scaling up spin-based detection systems. Furthermore, photoelectrical charge carrier generation does not saturate at high excitation powers \cite{bourgeois_photoelectric_2015}, enabling higher laser intensities and thus improved readout fidelity compared to optical measurements \cite{nishikawa_coherent_2025, hrubesch_efficient_2017}.\\
\indent Electrical spin readout has been demonstrated for a wide range of materials such as diamond for singles \cite{siyushev_photoelectrical_2019} and ensembles \cite{bourgeois_photoelectric_2015} of nitrogen vacancies, hexagonal boron nitride for negatively charged boron vacancies \cite{ru_room-temperature_2025}, gallium arsenide quantum dots \cite{koppens_driven_2006} and phosphorus donor electrons in silicon \cite{stegner_electrical_2006}. However, PDMR in SiC has only been demonstrated for bulk experiments with a single \cite{nishikawa_coherent_2025} and an ensemble \cite{niethammer_coherent_2019, nishikawa_electrical_2022} of silicon vacancies including a SiC n$^+$-p junction diode \cite{lew_all-electrical_2024-1} and metal–oxide–semiconductor field effect transistor (MOSFET) \cite{abe_electrical_2022}. Implementing PDMR in a platform that supports integrated photonics is a step toward scalable spin initialisation and detection systems, reducing the need for bulk optics and photodetectors. \\
\indent SiC integrated photonics require a thin SiC layer on the order of several microns, bonded to an insulating substrate—such as silicon dioxide (SiO$_2$). This combination is referred to as silicon carbide on insulator (SiCOI). Photonics in SiCOI  have already demonstrated low-loss optical devices \cite{lipton_low-loss_2025,wang_high-q_2021}, ultrahigh-Q SiC photonic crystal nanocavities \cite{song_ultrahigh-q_2019} fast optical modulators \cite{powell_integrated_2022} and integration with defects \cite{hu_room-temperature_2024}, making SiCOI a promising platform for scalable quantum technologies \cite{bader_analysis_2024}. Implementing PDMR on a thin film SiCOI platform thus provides an elegant solution to scale up hundreds of on-chip spin state detection systems without the need for collection optics. Combining PDMR and SiCOI would be particularly useful in applications where only local readout and no photon entanglement is required, such as waveguide based quantum sensing \cite{fisher_high-resolution_2025} and spin-spin entangled computation \cite{burkard_spin-entangled_2007}. Furthermore, waveguides enhance light–matter interactions by efficiently increasing the interaction volume, thereby enabling efficient readout of PDMR, reducing the required laser power, and making it more compatible for biosensing applications. \\
\indent In this work, we demonstrate for the first time room temperature PDMR of a small ensemble ($\sim$ 540) of V2 silicon vacancies in a 4H-SiCOI platform. Furthermore, we compare optical and electrical readouts in bulk and thin-film SiC, establishing that our fabrication process has no significant effect on the PDMR signal and $T_2
$ coherence time. Finally, we report electrical spin state readout across wavelengths ranging from 780 nm to 990 nm. Demonstrating for the first time electrical spin state readout well beyond the zero phonon optical transition energy (ZPL) of the V2 $\mathrm{V}_{\mathrm{Si}}^{-}$ and having optimal contrast within a wide range of wavelengths around the ZPL $\pm$ 20 nm. \\
\newpage

\begin{figure}[h!]
\begin{center}
    \includegraphics[scale=0.4]{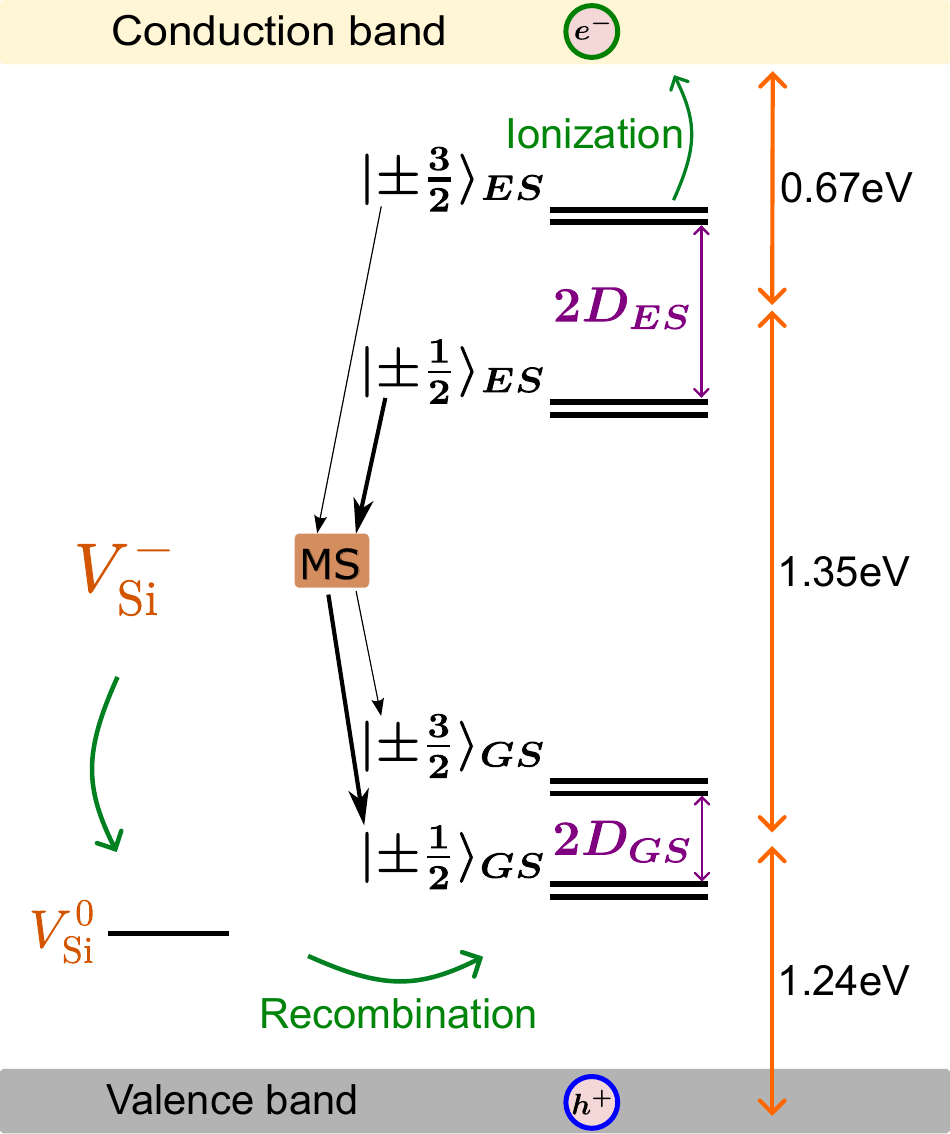}
  \caption[]{Spin quartet energy states of the  $\mathrm{V}_{\mathrm{Si}}^{-}$ with the optical transition of 1.35 eV between ground and excited state of the $\mathrm{V}_{\mathrm{Si}}^{-}$. Green arrows indicate the charge cycling. Photon-induced photocurrent from the defect is enabled by  electron generation in the conduction band via two-photon absorption and photon-induced hole generation in the valence band. }
  \label{fig:Vacancy_States}
\end{center}
\end{figure}
\noindent{\bfseries The V2 silicon vacancy and detection}\\
\indent The silicon vacancy at the cubic lattice site $\mathrm{V}_{\mathrm{Si}}^{-}$, consists of five active electrons: four from dangling bonds surrounding the vacancy and one captured electron. This leads to an optically active ground state with spin quartet configuration (S=3/2) \cite{mizuochi_continuous-wave_2002}. The zero phonon optical transition energy (ZPL) of the V2 $\mathrm{V}_{\mathrm{Si}}^{-}$ is 1.35 eV (917 nm) \cite{baranov_silicon_2011}. Off-resonant laser excitation promotes the $\pm \ket{3/2}$ and $\pm \ket{1/2}$ spin states to the excited state (Figure \ref{fig:Vacancy_States}). Relaxation occurs either through non-radiative decay via an inter-system crossing (ISC) into a metastable state (MS), or via spin-conserving radiative decay, yielding room-temperature broadband phonon assisted emission from 850 nm to 1000 nm \cite{fuchs_engineering_2015}. Off-resonant laser excitation pumps the $\pm \ket{1/2}$ state which preferentially decays through the ISC (dark), while the $\pm \ket{3/2}$ decays radiatively (bright). These spin sub-levels are separated by a zero-field splitting of $2{D_{gs_{0}}} \approx $ 70 MHz. Through simultaneous laser illumination and resonant RF driving at $2{D_{gs_{0}}}$, an optical spin contrast is observed, defined as a percentage (\%) given by
\begin{equation}
        \left(\frac{S_{RF}}{S_0} - 1\right)\times 100
\end{equation}
  where $S_{RF}$ and $S_0$ is the detected signal magnitude with and without RF drive respectively. In the case for optical detection, $S$ is the measured count-rate.

\indent An external magnetic field $B_0$  along the SiC crystal c-axis lifts the spin state degeneracy, providing a field-dependent Zeeman splitting given by  $2D_{gs} =  2{D_{gs_{0}}}\pm \gamma B_0$  where $\gamma$ is the electron gyromagnetic ratio. By tuning the RF frequency to $2D_{gs}$, one can address the individual spin states of the defect. \\
\indent Electrical readout of the spin-state utilises the technique of PDMR. This is possible by first exciting the electrons within the $\mathrm{V}_{\mathrm{Si}}^{-}$ to the excited state (1.35 eV), and a second photon ionises the $\mathrm{V}_{\mathrm{Si}}^{-}$ to the neutral charge state $\mathrm{V}_{\mathrm{Si}}^{0}$ promoting an electron to the conduction band (CB) \cite{nishikawa_coherent_2025}. Recently it has been proposed that an ionisation path through the doubly negative charge state $\mathrm{V}_{\mathrm{Si}}^{2-}$ might be a possible mechanism for electrical detection \cite{steidl_single_2025}. Finally, absorption of a third photon generates a hole in the valence band (VB). This charge cycling process through the excited state \cite{nishikawa_coherent_2025} forms the basis for spin dependent photocurrent (Figure \ref{fig:Vacancy_States}).\\
\indent An electrical spin state contrast is possible by driving an RF frequency matching $2D_{gs}$, increasing the $\pm \ket{3/2}$ population. This inherently decreases the metastable state occupation, and therefore more electrons are available for ionisation to the conduction band, leading to a measurable increase in photocurrent. Electrically detected spin state contrast is calculated using equation (1), the same as optical measurements, however signal $S$ is the magnitude of the measured photocurrent.\\ \indent Electrical readout also collects charge contributions from other defects and free carriers.  Therefore, the total detected photocurrent under laser illumination is the sum of all charge contributions, including from the $\mathrm{V}_{\mathrm{Si}}^{-}$, and is referred to as the laser-induced background. \\

\begin{figure}[h!]
\begin{center}
    \includegraphics[scale=0.41]{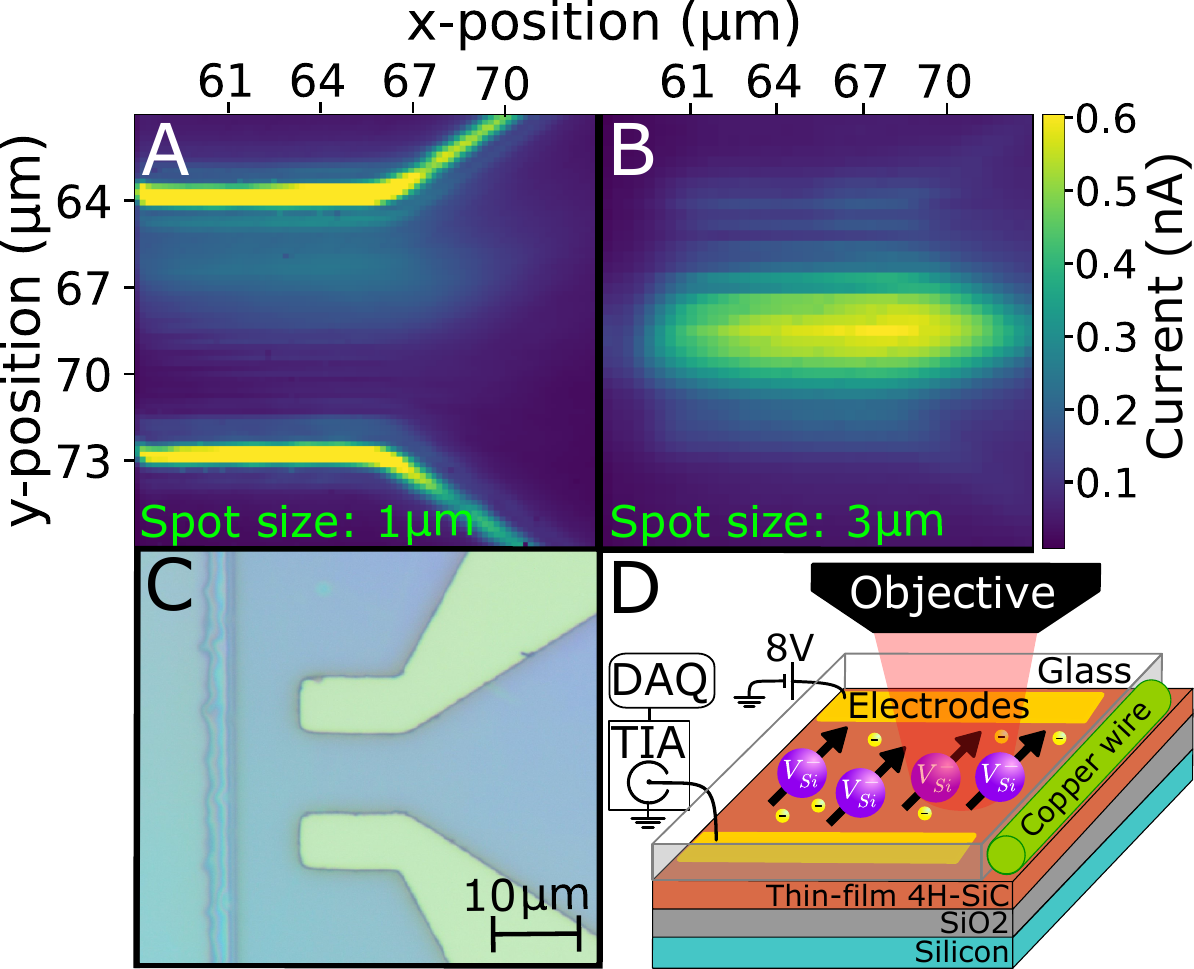}
  \caption[]{Confocal photocurrent scans (A and B) of the 4H-SiCOI device (D) at 23 mW, 10 V bias, and 870 nm. The 1µm spot size (A) has a maximum photocurrent of 1.2 nA at the edge of the electrodes, and the 3 µm spot size (B) has a maximum photocurrent of 0.8 nA in-between the electrodes. The larger spot size demonstrates a broader distribution of photocurrent extraction. A microscope photograph (C) of the electrode device covered in HSQ.}
  \label{fig:FullDevice}
\end{center}
\end{figure}

\noindent{\bfseries SiCOI device and measurement scheme} \\
\indent Commercially available research grade high-purity-semi-insulating (HPSI) 4H-SiC was bonded to a SiO$_2$ insulator with silicon as a substrate. The bonded SiC was ground, polished and etched down to a thickness of 1.3 µm, the three layer device is shown in Figure \ref{fig:FullDevice}D.  Gold electrodes of 150 nm with a 10 nm chromium adhesion layer were patterned on the SiC surface using standard optical laser lithography. The electrodes were covered with optically transparent hydrogen silsesquioxane (HSQ), to mitigate environmental fouling by moisture and airborne particulates, leaving patterned windows only for bonding pads and the 50 µm enamelled copper RF line shown in Figure \ref{fig:FullDevice}C. The gold electrodes were wire bonded with aluminium wire to a 50 $\Omega$ impedance matched printed circuit board and biased with a linear power supply for charge extraction. The photocurrent was measured using a 1 kHz $10^9$ V/A transimpedance amplifier (TIA) directly connected to a 2 Ms/s analog data acquisition device (DAQ). The experimental ground was an isolated unanodized aluminium breadboard directly connected to the building ground in which all electronic devices shared. This configuration measured a dark current of 80 pA and measurement noise of 20 fA with a bias of 8 V. Zeeman energy level splitting was induced with a permanent magnet placed underneath the sample, and all measurements were performed at room temperature.\\
\indent Optical excitation was performed on the surface between the electrodes using a 662 to 1050 nm wavelength selective Ti:Sa CW Ring Laser. By lowering the focal position of the objective, the laser-spot size on the SiC surface is increased, resulting in different current extraction regimes. An electrical confocal image was constructed by scanning across a 10 µm $\times$ 10 µm region in 300 nm steps using 23 mW of 890 nm pump laser light with 10 V biased electrodes and measuring the photocurrent at each point for 200 ms shown in parts A and B of Figure \ref{fig:FullDevice} for spot diameters of 1 µm and 3 µm, respectively. Figure \ref{fig:FullDevice}A shows  a maximum photocurrent of 1.2 nA along the electrode edges. Figure \ref{fig:FullDevice}B is defocused into the sample, demonstrating a larger, more uniform distribution of charge collection with a maximum photocurrent of 0.8 nA in-between the electrodes, and less variance in the photocurrent (see Figure \hyperref[sec:supp]{S4}). \\
\indent The number of measured silicon vacancies in each focus regime was estimated by considering the photon emission rate of 3-4 kc/s for a single V2 $\mathrm{V}_{\mathrm{Si}}^{-}$ measured in an identical optical collection setup used in ref \cite{fisher_high-resolution_2025} with a 805 nm long-pass dichroic. A 1 µm diameter laser spot focused on the SiCOI surface collected a maximum measured photon emission of 80 kc/s without HSQ cover and with a 900 nm long-pass, cutting approximately half the phonon emission sideband \cite{fuchs_engineering_2015} equating to $\sim$ 40 to 50 spins. A spot size (Airy diameter) of 3 µm equates to 9 times larger illumination area, therefore in the defocused regime we estimate an electrical signal comprised of approximately 540 $\mathrm{V}_{\mathrm{Si}}^{-}$'s including the possible contribution of higher order Airy rings (20 \%) generating a photocurrent. The estimation does not account for the possibility that more defects may be illuminated and may contribute to the photocurrent than are illuminated and collected in the optical signal.
\\
\indent Pulsed spin control measurements were performed by sending a continuous RF signal through a fast RF switch and amplifying the switched RF signal for the spin control. The AOM, RF-switch and DAQ readout were triggered with a GHz TTL pulse sequence streamer. Pulsed PDMR and Rabi sequences were continuously streamed for each frequency or RF duration with an 8 Hz square envelope modulating the RF amplitude for signal and reference measurements. The laser pulse duration was set to 1338 ns, with the MW pulse leading edge set at 1000 ns plus the maximum MW pulse duration, after each laser pulse leading edge, varying the duration of the MW pulse by extending the trailing edge and keeping the time between laser pulses constant, ensuring a consistent integer number of laser pulses for each envelope period. The analog voltage from the TIA output was collected every 500 ns by DAQ, integrating a window within the 8 Hz envelope half period for the highest signal to noise ratio (SNR). For optical measurements, laser and emission were filtered with a 900 nm long-pass dichroic and further filtered with a long pass filter before collection with a fiber-coupled avalanche photodiode (APD). Photon counts were collected for a 300 ns window at the beginning of each laser pulse, ensuring that the signal is proportional to the state population, with a delay accounting for AOM rise time.  \\

\noindent{\bfseries Comparison of optical and electrical spin readout in SiCOI and bulk}\\
\begin{figure*}[h!] 
\begin{center}
    \includegraphics[scale=0.87]{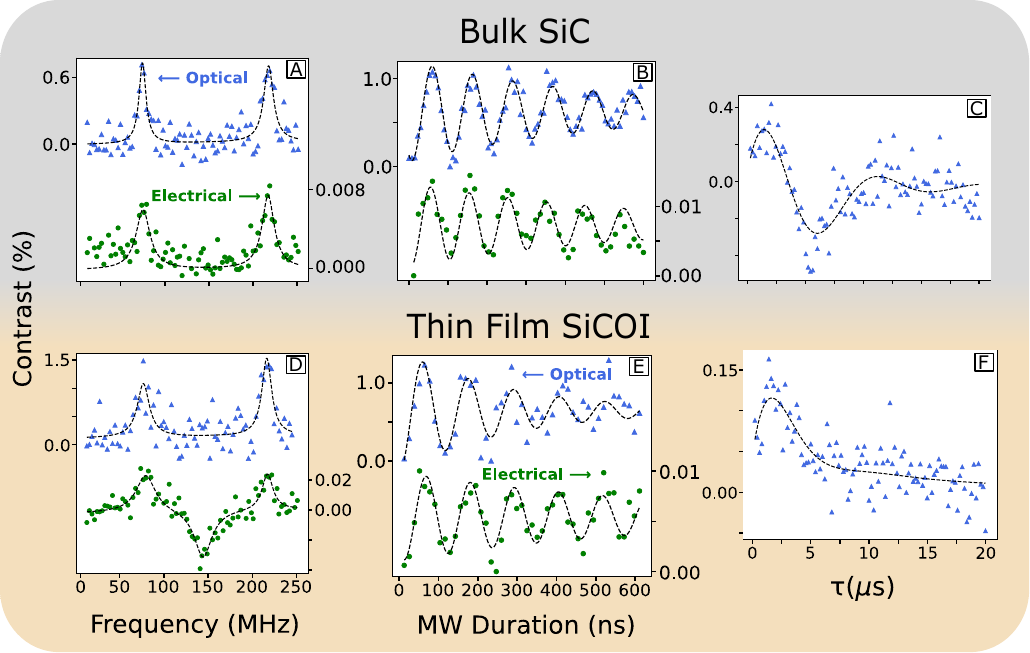}  
  \caption[]{Measured pulsed PDMR and pulsed ODMR in a 1.3 µm thin-film SiCOI device (D) and bulk 500 µm sample (A). All electrical measurements performed at 8 V bias, 890 nm and 45 mW. Electrical Rabi oscillations are observed with contrast $\approx$ 0.01\% for both bulk (B) and thin film (E). Optical $T_2$ Hahn echo measurement fitted with an ESEEM function (C and F) demonstrating coherence times of $\approx$ 7 $\pm 0.5$ µs for the thin film and bulk.}
  \label{fig:Rabi_PDMR_T2_Combined}
\end{center}
\end{figure*} 
\indent Pulsed PDMR was successfully demonstrated in the 4H-SiCOI platform (Figure \ref{fig:Rabi_PDMR_T2_Combined}D) and in a 500 µm bulk sample (Figure \ref{fig:Rabi_PDMR_T2_Combined}A) from the same original 4H-SiC wafer, with identical electrodes at 890 nm, 45 mW and 8 V bias and covered with the same transparent layer of HSQ. The bulk electrical PDMR has a contrast of 0.008\% and the thin film contrast is 0.025\% calculated using equation (1). PDMR results are fitted to two Lorentzian functions, three for the case of the negative dip in the SiCOI measurement which has been observed in previous studies and possibly due to excitation of the V1' silicon vacancies with minimal change to the relative V2 signal \cite{fisher_high-resolution_2025, nagy_quantum_2018}.   \\
\indent Electrical Rabi oscillations were also observed for both bulk (Figure \ref{fig:Rabi_PDMR_T2_Combined}B) and thin film (Figure \ref{fig:Rabi_PDMR_T2_Combined}E) under the same measurement conditions, with a contrast of 0.01\%, confirming  PDMR readout of silicon vacancies within the SiCOI layer after grinding, polishing and inductively coupled plasma (ICP) etching down to 1.3 µm (suitable for waveguides). The corresponding Rabi oscillations were fitted to exponentially damped sinusoidal functions using least squares regression. Laser powers from 30 to 50 mW showed no measurable change in electrical Rabi contrast. The measured contrast was limited by the density of spin ensemble and background photocurrent from non-$\mathrm{V}_{\mathrm{Si}}^{-}$. The highest reported PDMR contrast can be up to 0.4\% for single silicon vacancies \cite{nishikawa_coherent_2025} with reduction of background photocurrent from partially insulating electrodes. Furthermore, SiCOI demonstrated a small decrease in background photocurrent compared to the bulk (Figure \hyperref[sec:supp]{S1}). Contrast is further dependent on the laser position inside the electrodes and required positional stabilisation for consistent results (Figure \hyperref[sec:supp]{S3}). \\
\indent For optical measurements, the laser was under fully focused conditions, maximising photon counts for both the thin-film and bulk, locked to an ensemble inside the electrodes within the same approximate position as electrical measurements.  A laser power of 2 $\pm$0.5 mW was used, which was close to the saturation point of the defects. The measured optical Rabi contrast is around 1\% for the thin-film and bulk.   \\
\indent Electrical readout exhibited electrical spin contrast 75 times lower than optical spin contrast. However, electrical measurements had a higher SNR than optical measurements from 840 nm (Figure \hyperref[sec:supp]{S6}) due to poor optical collection efficiency. Disregarding contrast and laser noise, the measured optical SNR was limited by photon collection efficiency, photons from other defects and APD detector efficiency and noise. The electrical SNR was limited by noise from laser-induced background. Photocurrent measurements were not yet limited by TIA noise or noise in the linear power supply bias voltage because tests with a solid-state battery ($\pm <$1 µV) showed no reduction in photocurrent variance. Therefore, electrical spin contrast measurements were limited only by electron shot noise from the SiC charge extraction and laser power fluctuations. \\
\indent Regarding signal improvement (increasing the SNR), electrical readout studies with different bias voltages have been reported \cite{niethammer_coherent_2019} and demonstrate a small increase in SNR for higher voltages. However our goal is to perform electrical readout using CMOS compatible voltages (< 5 V). We thus look toward reducing electrode sizes and spacing to reduce the background photocurrent and required voltage, ensuring that the electrode geometries match the illuminated defect region. This reduces the leakage current from nonilluminated regions. Another prospect for contrast improvements toward single-defect PDMR involves reducing the background with partial electrode insulation, as demonstrated in ref \cite{nishikawa_coherent_2025}.\\
\indent Hahn echo decay was measured and fit with an electron spin-echo envelope modulation (ESEEM) function, showing optical spin coherence times of $\sim$ 7 $\pm$ 0.5 µs for both the thin film and bulk SiC, which demonstrates a similar $T_2$ coherence time of the defects in both the bulk and fabricated thin-film SiC on insulator after grinding, polishing, and ICP etching. A fast Fourier transform of the raw Hahn echo decay data reveals frequency components around 150-200 kHz for thin-film and bulk samples (Figure \hyperref[sec:supp]{S7}). The magnetic field strength of the permanent magnet near the carbide is  $B_0 \approx $ 5 mT, calculated from the Zeeman energy level splitting in the ODMR data. Electrical measurements of the Hahn echo decay required 10 µs between laser pulses to keep the number of pulses fixed in the modulated RF amplitude envelope while RF pulse spacings were swept. This made measurements greatly susceptible to laser fluctuations. Measurement attempts were made, however, due to low SNR from laser power instabilities, electrical Hahn echo decay results were inconclusive.\\

\noindent{\bfseries Wavelength dependence of electrical and optical Rabi in SiCOI}\\ 
\begin{figure*}[h!] 
\begin{center}
    \includegraphics[scale=0.85]{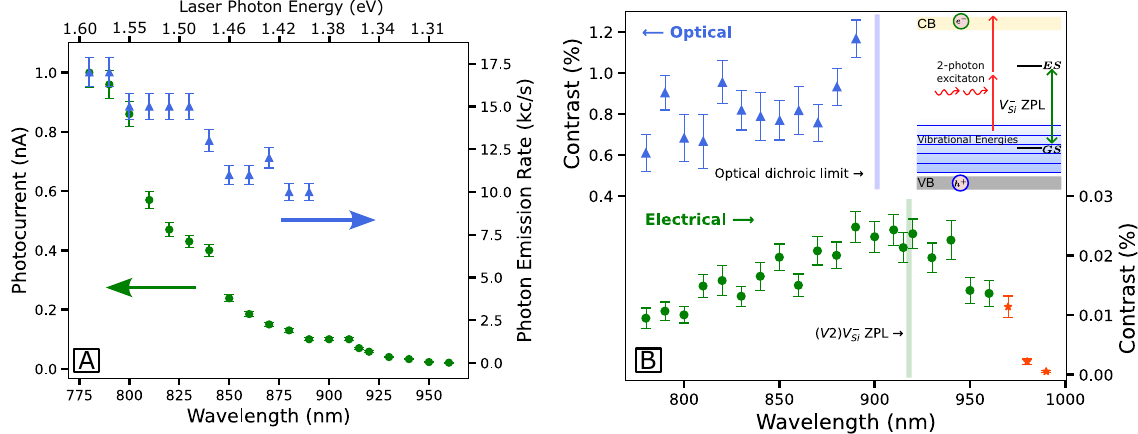}
  \caption[]{Wavelength dependence of the photocurrent and photon counts  (A), and Rabi contrast (B) in 1.3 µm thin-film 4H-SIC on SiO$_2$. Each data point in part B is one Rabi period, fitted to a decaying sine function. Rabi is measured electrically with 40 $\pm 5$ mW to 960 nm (green circles) and 40 to 70 mW up to 990 nm (red stars) with the respective photocurrents shown in (A). Photon counts correspond to the optical measurements (blue triangles) performed at 2 $\pm$ 0.5 mW. The optical error is directly from the fit, and the electrical error is the 5\% variation from experimental stability combined with the fit error. The top-right inset in part B schematically demonstrates the two-photon ionisation (red arrows) and phonon-assisted excitation (antistokes), enabling spin contrast beyond the ZPL optical transition.}
  \label{fig:Contrast_Wavelength}
\end{center}
\end{figure*}
\indent Previously, laser wavelengths of 785 nm \cite{niethammer_coherent_2019} and 905 nm \cite{nishikawa_coherent_2025, nishikawa_electrical_2022} have demonstrated electrical readout of the $\mathrm{V}_{\mathrm{Si}}^{-}$ spins in silicon carbide. In this work, the excitation wavelength dependence of the Rabi contrast was investigated from 780 nm to 990 nm electrically and 780 to 890 nm optically, with results provided in Figure \ref{fig:Contrast_Wavelength}B. Optical contrast shows a slight increase from 850 nm, consistent with observations from other optical-contrast wavelength studies of the $\mathrm{V}_{\mathrm{Si}}^-$ in SiC \cite{carter_spin_2015}. Electrical contrast increases from 780 nm to about 900 nm likely due to reduction in non-$\mathrm{V}_{\mathrm{Si}}^-$ electrons contributing to the laser-induced background (see Figure \ref{fig:Contrast_Wavelength}A).

It is reported that in some HPSI types (depending on the supplier), the $\mathrm{V}_{\mathrm{Si}}^-$, carbon vacancy ($\mathrm{V}_{\mathrm{C}}$), and C antisite–vacancy pair ($\mathrm{C}_{\mathrm{Si}}\mathrm{V}_{\mathrm{C}}$) are dominant \cite{son_charge_2021}. The energies required for ionisation and excitation of the $\mathrm{V}_{\mathrm{C}}$ are 0.66 and 0.8 eV respectively \cite{son_charge_2021}, enabling a contribution to the background photocurrent for this measured wavelength range. $\mathrm{C}_{\mathrm{Si}}\mathrm{V}_{\mathrm{C}}$ requires energies ranging from 0.72 to 1.07 eV for ionisation \cite{nakane_deep_2021}. $\mathrm{C}_{\mathrm{Si}}\mathrm{V}_{\mathrm{C}}$ exhibits calculated ZPLs at 1.16 and 1.33 eV with a strong absorption line at 1.56 eV \cite{bulancea-lindvall_temperature_2024}. This hypothetically explains the reduction in the defect-contributed background from 780 nm onward (Figure \ref{fig:Contrast_Wavelength}A). The wavelength dependent background photocurrent depends on the defects present in SiC. To reduce the contribution of other defects to the unwanted background photocurrent, it is important to consider the type of SiC being used, what defects may be intrinsically present, and how annealing techniques can remove or create them. Thus, the contrast of the spin system of interest may be improved.

The spin contrast exhibits a positional dependence, likely arising from spatial inhomogeneities in the defect distribution within SiCOI, where the varying distribution of $\mathrm{V}_{\mathrm{Si}}^-$  and non-$\mathrm{V}_{\mathrm{Si}}^-$ defects leads to regions with differing electron extraction efficiencies. A maximum-photocurrent optimiser was employed to stabilise the experiment against positional drift and produce experimentally repeatable results; however, maximising the photocurrent does not necessarily correspond to the region of highest contrast. Considering wavelength dependent charge extraction, shorter wavelengths are more likely to reach one-photon ionisation thresholds of certain defects, enhancing their contribution to the photocurrent, while longer wavelengths produce larger spot sizes (e.g., a 31\% increase in area at 990 nm compared to 780 nm). From Figure \hyperref[sec:supp]{S1}, the photocurrent response to a 31\% change in spotsize area within the stable measurement region is 37 \%. Consequently, the optimiser may converge on different spatial regions for different wavelengths, meaning that the observed wavelength-dependent trends can include variations in the probed sample volume corresponding to the maximum photocurrent. Nevertheless, these effects may be minimal when also considering that the Rabi rate fluctuated by 6\% across all wavelengths and the photocurrent reduction is much greater across the wavelengths (1 nA to 0.08 nA shown in Figure \ref{fig:Contrast_Wavelength}A). 

Excitation and photon collection near the ZPL is challenging due to the requirement for spectral filtering. On the other hand, electrical detection does not require collection optics. Thus, a clear spectrum of the electrical Rabi contrast was observed up to 980 nm, with a small signal at 990 nm. This demonstrates for the first time electrical readout of the spin state well beyond the ZPL of the $\mathrm{V}_{\mathrm{Si}}^{-}$. We hypothesise that at room temperature this process is enabled via phonon assisted excitation of the electron from the ground state to the excited state of $\mathrm{V}_{\mathrm{Si}}^{-}$ \cite{shang_local_2020, wang_robust_2021}. Previous studies have demonstrated phonon energies persisting from 1.35 to 1.2 eV for V2 $\mathrm{V}_{\mathrm{Si}}^-$ \cite{udvarhelyi_vibronic_2020}, consistent with the demonstrated spin contrast beyond the ZPL. Furthermore, spectral dependence and observation of the electrical Rabi contrast is demonstrated beyond the ZPL due to the threshold energy for ionisation and recombination  being lower than the defect excitation energy \cite{steidl_single_2025} combined with a band of antistokes vibrational energy contributions \cite{wang_robust_2021} potentially further assisting the ionisation process of $\mathrm{V}_{\mathrm{Si}}^-$ (Figure \ref{fig:Contrast_Wavelength}B, inset). The significant contrast reduction from 940 nm onward, is likely due to a reduction in the excitation of the $\mathrm{V}_{\mathrm{Si}}^-$ and ionisation efficiency, possibly explained by previous works measuring the two-colour excited photocurrent of $\mathrm{V}_{\mathrm{Si}}^-$ \cite{steidl_single_2025} demonstrating a threshold of 948 nm for the $\mathrm{V}_{\mathrm{Si}}^-$ to $\mathrm{V}_{\mathrm{Si}}^{2-}$ conversion. 

Electrical detection provides a wavelength-tolerant readout of the $\mathrm{V}_{\mathrm{Si}}^{-}$ spin state, not limited by wavelength dependent photon collection optics and devices. Spin-dependent signals were observed electrically for all tested wavelengths. Further investigations to wavelength dependent spin state processes can be undertaken using electrical readout. \\

\noindent{\bfseries Conclusion}\\ 
\indent In summary, we have demonstrated electrical spin readout and coherent control of silicon vacancies in 4H-SiCOI over a wide range of wavelengths, establishing a route toward scalable, CMOS-compatible quantum devices in SiC. By implementing PDMR in a 1.3 µm thin-film SiCOI architecture and comparing bulk SiC samples from the same wafer, we measure similar $T_2$ coherence times despite undergoing bonding, polishing and ICP etching, observing clear electrical Rabi oscillations and optical Hahn-echo coherence of $\sim$ 7 $\pm$ 0.5 µs in both the thin-film and unprocessed bulk material. \\
\indent Furthermore, we show for the first time that PDMR contrast persists across a wide excitation range (780 to 990 nm). Excitation wavelengths from  780 nm onward are hypothesised to suppress a background photocurrent from unintended charge sources, improving the contrast and SNR. Wavelengths between 840 and 890 nm demonstrated a Rabi SNR higher than optical measurements due to poor collection efficiency (Figure \hyperref[sec:supp]{S6}). The contrast extends well beyond the 917 nm V2 $\mathrm{V}_{\mathrm{Si}}^{-}$ ZPL and having optimal contrast around the ZPL $\pm$ 20 nm, a regime difficult to access with conventional optical setups. Clear Rabi contrast persists up to 980 nm, with evidence of contrast at 990 nm.  The measurement limitation from wavelengths of 970 nm onward is that the photocurrent signal approaches the dark current. Investigating PDMR at cryogenic temperatures could provide insight into using the electrical readout of the spin for measurement of phonon vibronic states \cite{udvarhelyi_vibronic_2020, shang_local_2020}. \\
\indent This work combines SiCOI, a platform suitable for large-scale integrated photonic fabrication, with a scalable electrical spin-state detection technique that does not require collection optics. Integrated photonics enables independent spin initialisation across a SiCOI wafer, obtainable with optical waveguides. However, on-chip optical readout of these spin states requires fabricated single-photon detectors and demands optical frequency filtering of the initialisation laser with a high extinction ratio. Electrical readout, on the other hand, offers a more attractive path for scaling up thousands of on-chip spin detection systems based on patterned metal electrodes. A thin-film platform is essential for guiding light in integrated photonic waveguides. Our results thus provide the first evidence that electrical readout of spin systems (PDMR) is preserved after aggressive thin-film fabrication, a step toward realising future compact and scalable spin state detection via electrical readout and waveguides. By combining an array of electrical spin readout electrodes with an integrated photonic network, one could realise a quantum processing chip, useful for distributed quantum computing, enabled by a network of quantum processing modules \cite{main_distributed_2025}. Furthermore, in applications where optical entanglement is not required, an array of spatially separated PDMR electrodes on SiCOI—as shown in this work—can function as individual quantum sensor regions. This pixel array functions as a quantum charge-coupled device, hypothetically yielding a versatile quantum imaging sensor that delivers spatially resolved measurements of electromagnetic fields and temperature, making it applicable for understanding the structure of organic compounds and imaging of electronic devices with submicrometre spatial resolution, extending beyond the limitations of wide-field optical microscopy setups \cite{briegel_optical_2025}. 

\bigskip
\noindent{\bfseries Author contributions}

A.Z. and P.F. designed and built the experimental setup. A.Z. measured the experimental data and programmed the Python measurement software. A.Z. and B.H. designed and fabricated the electrodes, B.H. fabricated the SiCOI. A.Z. and R.C. wrote the manuscript. N.M. assisted with writing and provided insightful experimental suggestions. All authors discussed the results and commented on the manuscript, R.C. supervised and conceived the project.\\

\newpage
\noindent{\bfseries Funding acknowledgements} 

This work was supported by Bundesministerium für Forschung, Technologie und Raumfahrt (Q-SiCk, 13N16296)\\

\noindent{\bfseries Acknowledgements} 

Device fabrication was made possible through the facilities and staff of the cleanrooms of both the Center of Microelectronics, Ulm University, and the Centre for Applied Quantum Technology, University of Stuttgart.\\

\noindent{\bfseries Notes}

Recently we became aware of the demonstration of wavelength dependent photocurrent measurements in bulk SiC with single V1 and V2 colour centers \cite{okajima_photoionization_2026}.\\

\noindent{\bfseries \hyperref[sec:supp]{Supporting Information}}

Spotsize calculation and experiment stabilisation,
Photocurrent characterisation, SNR comparison between optical and electrical, 
Hahn echo fast Fourier transform, Silicon carbide on insulator fabrication, 
Experimental hardware and setup

\printbibliography
\clearpage
\section*{Supporting Information}
\label{sec:supp}

\subfile{SupportingInformation}

\end{document}

%% file: SupportingInformation.tex
\setcounter{figure}{0}
\renewcommand{\figurename}{Figure}
\renewcommand{\thefigure}{S\arabic{figure}}
\maketitle
\begin{figure}[h!]
\begin{center}
    \includegraphics[trim={0cm 0 0 0cm}, scale=0.5]{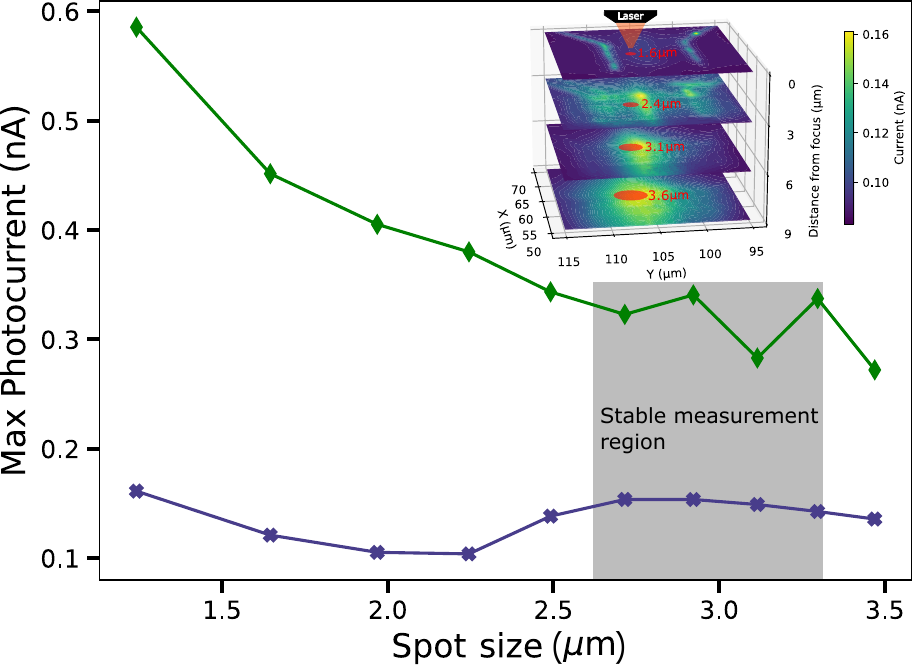}
  \caption[]{The maximum photocurrent taken from ten 20 µm $\times$ 20 µm confocal scans. The green diamonds are from the 500 µm bulk carbide and purple crosses from the 1.3 µm SiCOI sample each with $\Delta$z = 1 µm and plotted against estimated spot size. Four SiCOI electrical scans are shown in the inset with the spot size as reference and the grey shaded region indicates the spot size for a stable max photocurrent optimiser during measurements.}
  \label{fig:PC_Spotsize}
\end{center}
\end{figure}

\begin{figure}[]
\begin{center}
    \includegraphics[trim={0cm 0 0 0cm}, scale=0.48]{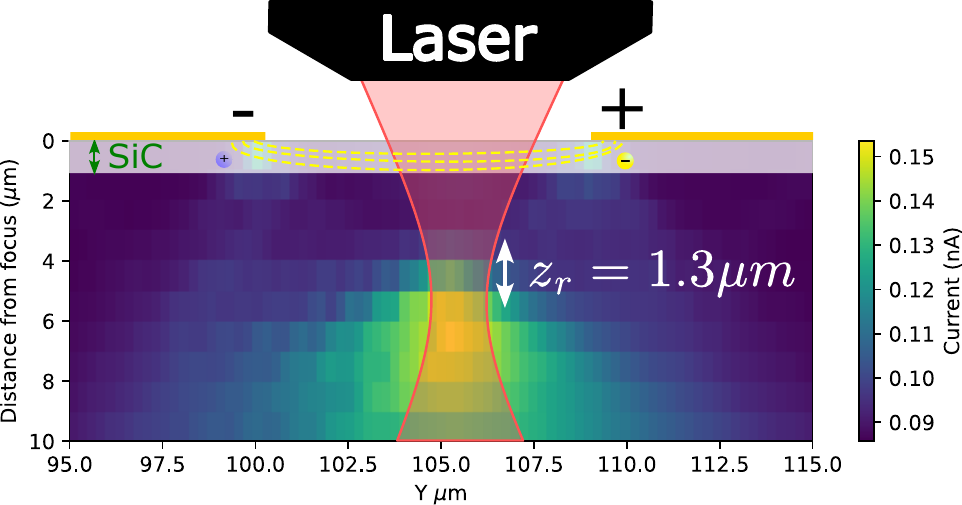}
  \caption[]{Electrical confocal scan along the beam propagation z-axis between electrodes in the 1.3 µm thin film on SiO$_2$ on silicon, measured at 890 nm, 8 V bias and 40 mW.  Charge collection is from the thin SiC layer on top. The Rayleigh length is annotated for reference, roughly scaled to the scan region dimensions. Shown is the focal configuration for the data in main text, with an optimiser scanning the sample for maximum photocurrent.}
  \label{fig:Z-Y_Scan}
\end{center}
\end{figure}

\begin{figure}[]
\begin{center}
    \includegraphics[scale=0.45]{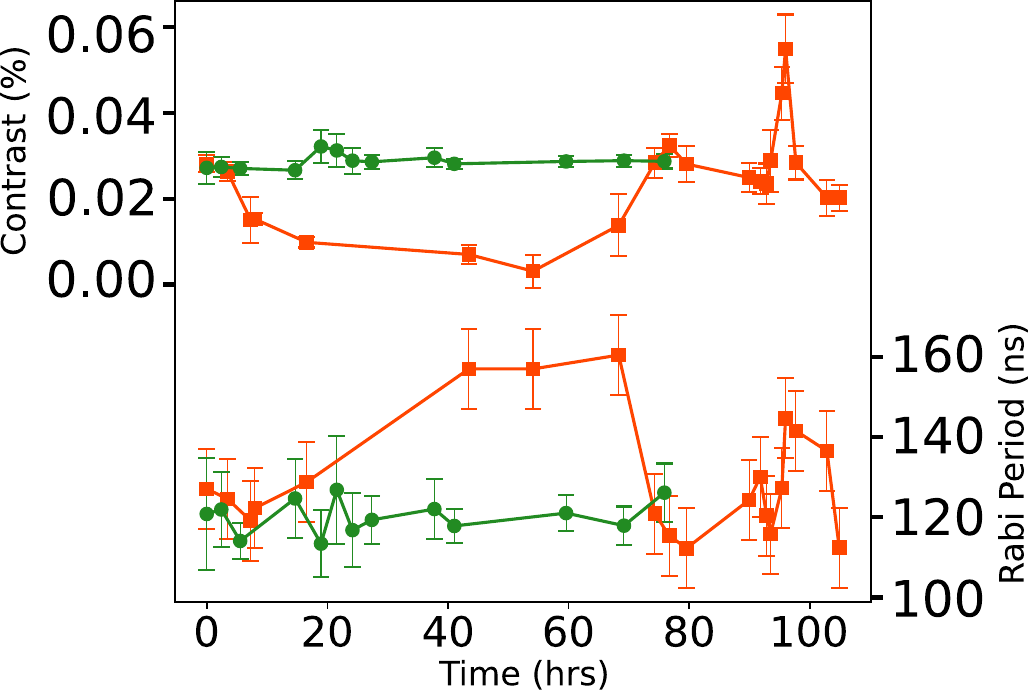}
  \caption[]{Contrast and Rabi period was monitored over 3 days with maximum PC optimiser (green circles) and 4 days with no optimiser (orange-red squares). Stabilised contrast standard deviation is $\approx$ 5\%  and combined with the fit error of a decaying sinusoidal function used in the main text wavelength measurement error bars.}
  \label{fig:Contrast_Stability}
\end{center}
\end{figure}

\begin{figure}[h!]
\begin{center}
    \includegraphics[scale=0.45]{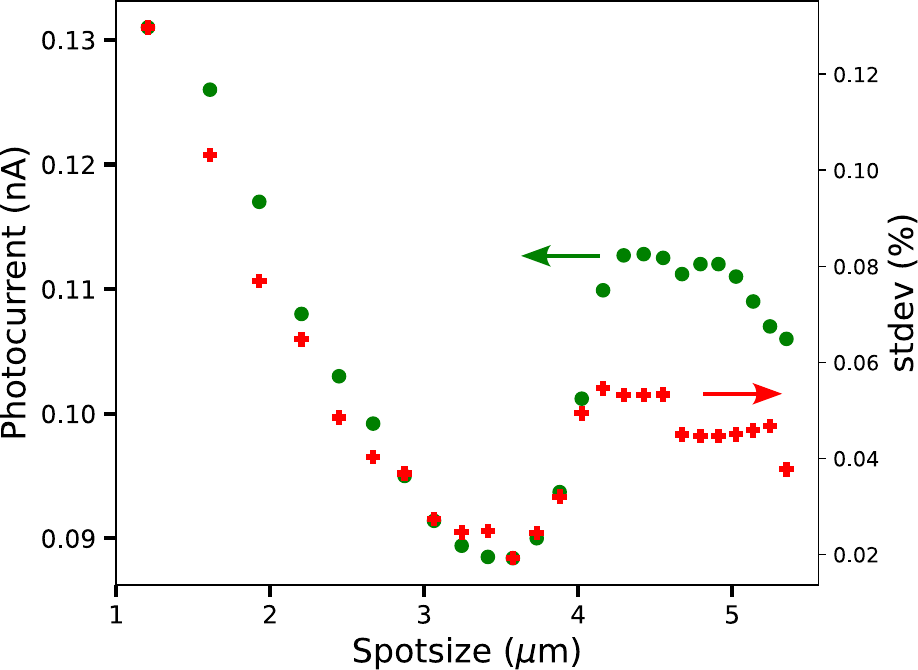}
  \caption[]{The standard deviation (red plus) and average (green circle) of fifty photocurrent measurements over 5 seconds plotted for each spot size at 890 nm, 37 mW and 8 V bias. Data taken from one fixed x-y position and beam propagation z-axis translated using a nano-positioning piezo stage. The relative standard deviation is shown to depend on the spot size.}
  \label{fig:PC_Spotsize_Var}
\end{center}
\end{figure}

\begin{figure}[]
\begin{center}
    \includegraphics[scale=0.45]{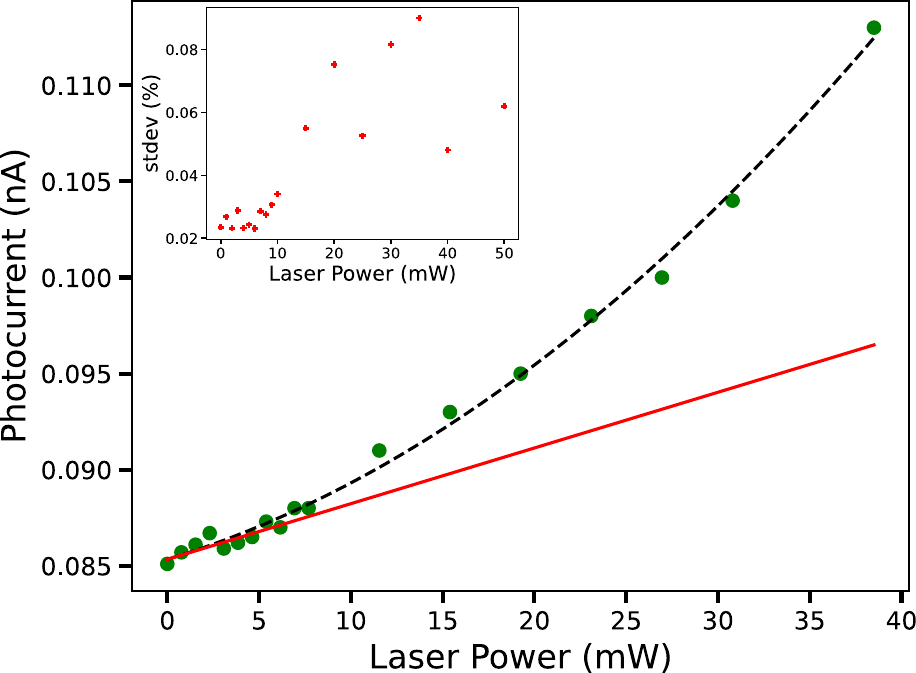}
  \caption[]{Photocurrent from the thin film using 890 nm and 8 V bias. The data is fitted to second order polynomial (dashed black line) and the linear component from the fit is plotted as a red solid line. The relative standard deviation shown in the inset is taken from 50 photocurrent measurements over five seconds. }
  \label{fig:PC_power}
\end{center}
\end{figure}

 
\noindent{\large \bfseries Spot size and stabilisation}\\
\indent We noticed that the photocurrent and contrast changes due to different spot sizes and positions in the sample; therefore, it was critical to find an optimal point for repeatable results. This specifically matters for the wavelength measurements, so they are not affected by changes in position or spot size. \\
\indent An objective Airy disk has a spot size measured by the radius from the centre to the first zero of the airy pattern intensity function given by $$\omega_0 = \frac{1.22\lambda}{2\text{NA}}$$
where $\lambda$ is the laser wavelength and NA is the numerical aperture of the objective. The spot size is a function of distance $z$ along the illumination direction given by $$\omega_z = \omega_0 \sqrt{1+\frac{z}{z_R}}$$
where $$z_R = \frac{\pi \omega_0^2}{\lambda}$$ is the the Rayleigh length of the beam. So therefore the beam-radius as a function of $z$ is 
$$\omega_z = \frac{1.22\lambda}{2\text{NA}}\sqrt{1+\frac{4\text{NA}^2z}{1.4884\pi\lambda}}$$
The smallest spot size is when the objective is in focus i.e, $z=0$. As the sample moves closer to the objective, the surface spot size is estimated by using the distance travelled by the piezo stage from focus. Measuring an electrical confocal scan for a  20 µm x 20 µm area and taking the maximum photocurrent (PC), the change in photocurrent per spot size is shown in Figure \ref{fig:PC_Spotsize}. 
We can see there is a peak photocurrent when the beam is in focus along the electrode edges, and another maximum in between the electrodes when the spot size is 2 to 3 µm.\\
\indent Performing a Z-Y scan and measuring the photocurrent, the photocurrent map along the z-direction is shown in Figure \ref{fig:Z-Y_Scan}. 
The calculated Rayleigh length for 890 nm is $z_r = $  1.3 µm and shown for reference. The top 1.3 µm is the silicon carbide (SiC) and source of charges. This configuration of focusing into the sample was the measurement condition for the data in the main text. \\
\indent With a larger spot size an optimiser can find the maximum photocurrent within a 7 µm $ \times $ 7 µm $\times$ 0.5 µm translation range every 5 minutes. To test stability, the contrast and Rabi rate was monitored over a period of days to ensure experimental repeatability in the contrast measurements. Figure \ref{fig:Contrast_Stability} shows the Rabi contrast and period over 3 days with maximum photocurrent optimiser and 4 days with piezo stage turned off (no optimiser). The maximum photocurrent optimisation algorithm at the stable point shown in Figure \ref{fig:PC_Spotsize} has a 5\% deviation in measured rabi contrast. With no optimiser the experiment naturally drifts due to temperature fluctuations, finding places with higher and lower contrast. Positional contrast dependence is likely due to having regions of higher $\mathrm{V}_{\mathrm{Si}}^{-}$ electron extraction since the distribution of defects is not completely uniform within the silicon carbide on insulator (SiCOI), and some regions could have more or less efficient charge extraction from the $\mathrm{V}_{\mathrm{Si}}^{-}$. One caveat of using the maximum current optimiser is that it does not find the region with maximum contrast; stabilisation to maximum contrast (not photocurrent) requires image processing algorithms to lock to a specific location within a confocal scan image and was not implemented in this work.\\
\indent Furthermore, by scanning the spot size at a fixed X-Y position, it was also discovered that the relative standard deviation in the photocurrent also changes. By taking 50 photocurrent measurements every 100 ms (five seconds total) for different spot sizes, the variance would decrease and increase along with the photocurrent average from the same 50 data points shown in Figure \ref{fig:PC_Spotsize_Var}. The variance was monitored before measurements, providing an indication of the measurement noise. \\
\newline
\noindent{\large \bfseries Photocurrent Characterisation}\\
\indent Photocurrent non-linearity was tested for a laser power range of 0 to 40 mW at 890 nm pump wavelength using an 8 V bias across the electrodes, demonstrating a small second-order quadratic relationship shown in Figure \ref{fig:PC_power}.
The linear component of the fit is plotted as a reference in red. The quadratic relation is a product of the two-photon ionisation process mentioned in the main text; however, this trend varies with position, voltage, and wavelength, most frequently showing a linear trend as most of the photocurrent signal is from non-$V_{\mathrm{Si}}^{-}$ electrons. The percent variance is also seen to increase with laser power, and plateaus from 20 mW with fluctuations possibly due to insufficient settling time of charge extraction from manually sweeping the laser power.  \\

\begin{figure}[h!]
\begin{center}
    \includegraphics[scale=0.5]{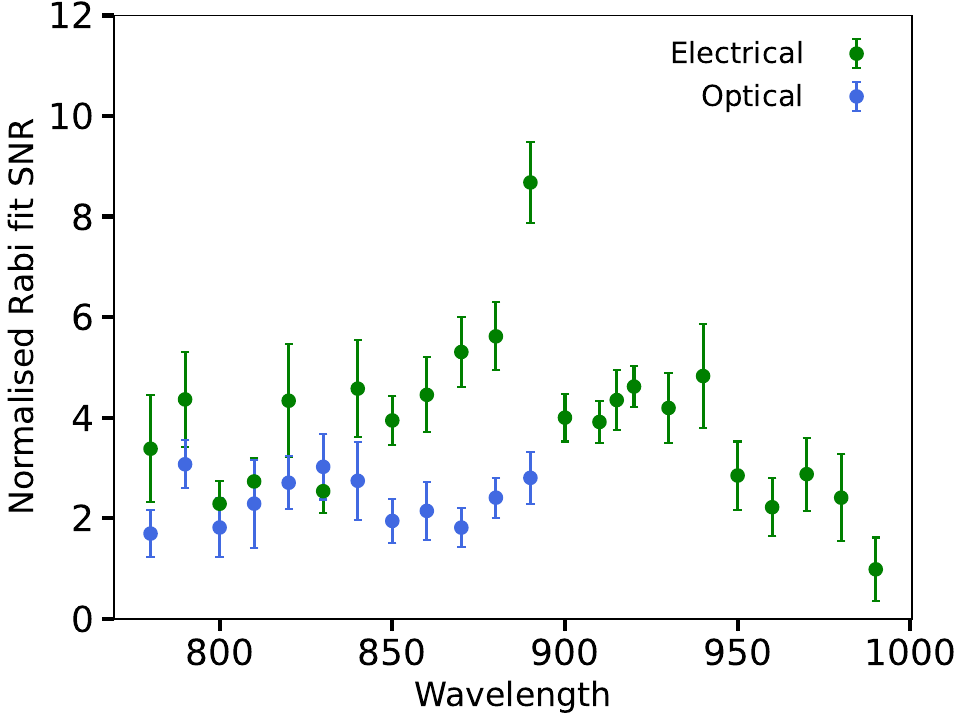}
  \caption[]{The signal-to-noise ratio (SNR) per wavelength, extracted from the Rabi oscillation fits in the baseline region, is defined as the ratio of the fitted Rabi contrast to the standard deviation of the residuals (data minus fit) in the wavelength scan shown in the main text. The optical collection efficiency was not optimal due to experimental limitations, which results in a lower SNR; nevertheless, these are the experimental conditions used throughout the main paper.}
  \label{fig:SNR_Wavlength}
\end{center}
\end{figure}

\indent For quantum technology applications, it is also important to check the SNR per wavelength for signal visibility and measurement time. By taking the ratio of the fitted contrast from the least-squares fit of an exponentially decaying sinusoid $C_{fit}$, and standard deviation of the residuals from the Rabi fit $\sigma$ (data minus fit), i.e. 
\begin{equation*}
   \mathrm{SNR}_{\mathrm{Rabi}} = C_{\mathrm{fit}}/\sigma
\end{equation*}
and then normalising with measurement time following the method outlined in \cite{niethammer_coherent_2019}, 
 \begin{equation*}
     \mathrm{SNR}_{\mathrm{norm}} = \mathrm{SNR_{\mathrm{Rabi}}} \times  \sqrt{3600/t} 
 \end{equation*}

where $t$ is measurement time in seconds), the Rabi SNR per wavelength is shown in Figure \ref{fig:SNR_Wavlength}. This method shows an increasing SNR up to 890 nm and consistent reduction from 940 nm for electrical measurements. Although this method of calculating the SNR is not using noise density analysis from the PDMR signal nor direct ratio of the $\mathrm{V}_{\mathrm{Si}}^{-}$ electrons and non-$\mathrm{V}_{\mathrm{Si}}^{-}$ electrons, it still indicates some wavelength dependent mechanism affecting the spin contrast noise. Further investigation could hypothetically provide insight into using electrical readout of the spin state to observe vibronic states within SiC \cite{udvarhelyi_vibronic_2020, shang_local_2020} and hypothetically wavelength-selective vibronic-assisted contrast enhancement or ionisation efficiencies. \\

\begin{figure}[h!]
\begin{center}
    \includegraphics[scale=0.5]{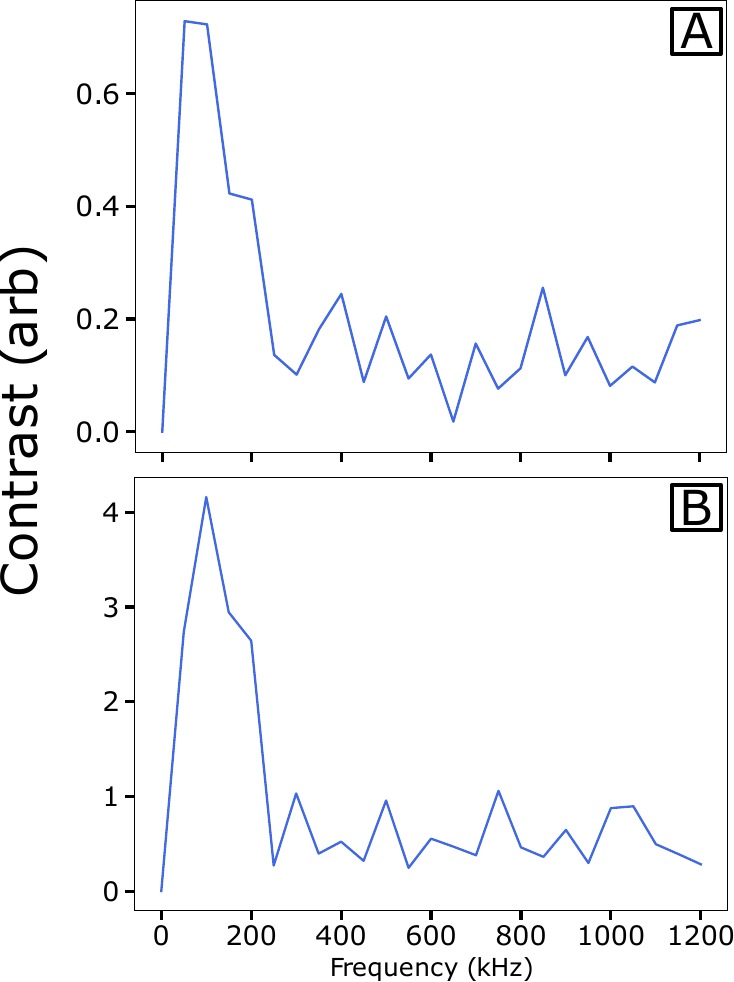}
  \caption[]{Fast Fourier Transform of the raw T2 Hahn echo decay for the thin-film (A) and bulk (B), demonstrating broad frequency components around 150 kHz to 200 kHz. The magnetic field strength at the defects is $\approx$ 5 mT.} 
  \label{fig:T2_FFT_Bulk_THinFilm}
\end{center}
\end{figure}

\noindent{\large \bfseries SiCOI and electrode fabrication}\\
\indent The SiC wafer is diced into 10 mm x 10 mm chips. A single side polished prime Si + 2000 nm dry/wet/dry SiO2 wafer, 4 inch, 525 µm thick, from Microchemicals GmbH is cleaved into 15 mm $\times$ 15 mm chips. Both the SiC and Si/SiO2 chips undergo a thorough cleaning procedure, with Piranha, RCA1, and RCA2 cleaning, followed by an oxygen plasma surface preparation. The chips are placed in a wafer bonder (SüSS SB6e), Under vacuum in the wafer bonder the chips are heated to 450$^\circ$ C and pressed together with a downward force of 1500 N for two hours. The chips are then cooled, the downward force released, and returned to atmospheric pressure. This bonded stack then undergoes grinding and polishing to thin the SiC to approximately 40 µm. This consists of a grinding step with a D80 diamond grinding wheel (Dopa Diamond Tools), followed by three polishing steps with successively finer diamond slurry (6 µm, 1.5 µm, 0.25 µm). Finally, the film undergoes dry etching with ICP-RIE to thin the film from 40 µm down to the final desired thickness of 1.3 µm.\\
\indent The sample was cleaned in piranha solution and electrode designs were patterned with a UV laser-lithography system and AZ10XT photoresist. After development, the sample underwent an oxygen plasma clean to remove surface contaminants. Chrome was deposited as an adhesion layer using electron beam evaporation with 10 nm thickness and 150 nm of Gold for conduction. Liftoff was possible after soaking for 2 hours in 80$^\circ$ C NEP then sonicated for 4 seconds. Allresist SX AR-N 8400, hydrogen silsesquioxane (HSQ) was spun to 1.4µm thickness, patterned leaving windows for bonding pads, and baked at 250$^\circ$ C for 30 mins to cure on top of the SiCOI surface. The sample was then glued and wire bonded with 25 µm aluminium wire to a 50 $\Omega$ impedance matched PCB. A 50 µm enamelled copper wire was strung perpendicular to the electrode tips and soldered to the same PCB. \\
\newline
\noindent{\large \bfseries Hahn Echo decay Fast Fourier Transform}\\
\indent The Fast Fourier Transform of the raw data from the T2 hahn echo decay is shown in \ref{fig:T2_FFT_Bulk_THinFilm}. The origin of such coupled frequencies to the electrons is unknown; however, it is hypothesised to originate from impurities within the bulk of the material, given the prominence of the signal. Provided the shape is not a consistent Lorentzian, multiple frequency components could be present. 

\noindent{\large \bfseries Experimental setup}\\
\indent The sample was mounted to a 3-axis piezo stage (Piezoconcept LT3-200) to move the sample into the focal positions of the microscope. Measurements were performed using a fixed path home-built confocal microscope equipped with a 0.9 NA air objective (Zeiss EC Epiplan-NEOFLUAR 100x/0.9 DIC) for laser focusing and fluorescence collection. Defect excitation was achieved using a fully automated Sirah Ti:Sa CW ring laser, modulated using an acousto-optic modulator (CASTECH CAOM-200-015-TEC-780-950-AF-A17). The collection path was filtered with two 900 nm longpass filters and one 1000 nm shortpass filter (Thorlabs FELH0900, FESH1000). Filtering the laser and collection paths using a 900 nm dichroic mirror (Thorlabs DMLP900). Fluorescence was focused
into a fiber coupled APD (Excelitas SPCM-AQRH-14-FC-ND) connected to a National Instruments DAQ (NI USB-6363) for data collection. \\
\indent One gold electrode was connected to the signal terminal of a  DLPCA-200 transimpendence amplifier (TIA) from Femto, set to 10 Hz low pass filter in DC coupling mode for an absolute current measurement. Connections were made using a standard coaxial cable for signal shielding and BNC to screw terminal to split the signal and ground terminals of the TIA. \\
\indent The second gold electrode was connected to either the positive or negative output terminal of a Siglent SPD3303 linear power supply for voltage biasing. Polarity was chosen by sweeping voltage ranges of 0 to 9 V while trying to maximise laser-induced photocurrent signal to dark signal ratio. The opposite power supply terminal was connected to an un-anodised aluminium breadboard, which served as a fully isolated ground plane sitting on a rectangle of rubber, on which the whole confocal system stood, this acted as the completely isolated ground plane. The ground plane was then directly connected to the building ground. This ground was shared with all measurement devices in the experiment, reducing the possibility of ground loops. \\
\indent The TIA ground terminal was directly connected to the breadboard ground plane, the metal casing of the uninsulated TIA sat on the breadboard shared also with the PCB and RF ground. EMF shielding was achieved by building a cardboard box wrapped in aluminium foil Faraday cage grounded to the same plane. \\
\indent RF pulses were delivered using a vector signal generator (Siglent SSG3021X-IQE), amplitude modulated with a fast switch (ZASW-2-50DRA+) and amplified

\begin{figure}[h!]
\begin{center}
    \includegraphics[scale=0.8]{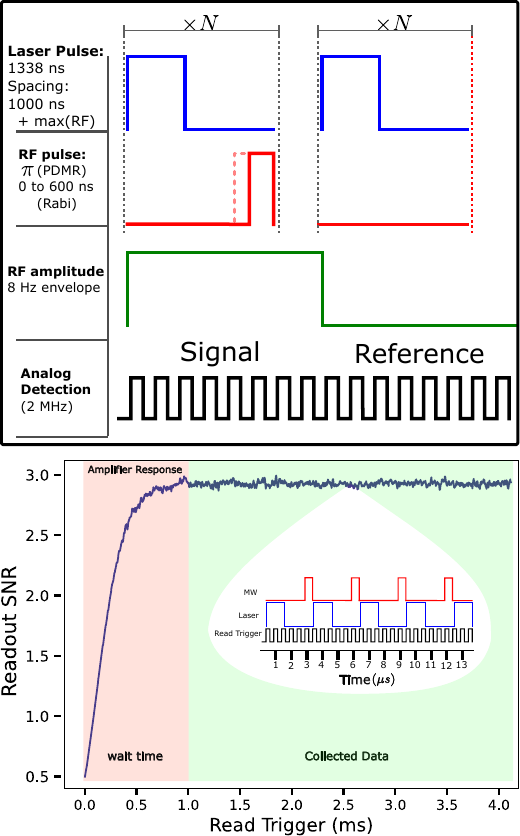}
  \caption[]{Schematic of the PDMR pulse sequence (top) with RF amplitude modulation envelope. Pulse sequence was repeated N times, where N = (0.5*envelopePeriod) / (laserDuration + RFSpacing + max(RFPulseDuration)), ensuring N is an integer. Data was integrated over each half envelope period for signal and reference. Bottom showing the wait time of the amplifier to account for the 1kHz response time then integrated over the repeating sequence while modulating the RF amplitude.}
  \label{fig:PulseSequence}
\end{center}
\end{figure}

(Mini-Circuits, ZHL-5W-1+). The AOM and RF pulses were synchronised using a Swabian Instruments Pulse Streamer. Rabi and pulsed PDMR sequences were streamed continuously for each randomised PDMR frequency (PDMR) or duration (Rabi), keeping the number of laser pulses the same integer number for all RF durations within an 8 Hz square wave period. This 8 Hz envelope was used for digital RF amplitude modulation signal and reference points. Analog voltage measurements were collected from the TIA directly using the DAQ, triggered at the sampling rate limit of the analog to digital converter (500 ns). Data from the first and second half of the 8 Hz envelope was programmatically separated into signal and reference (see Figure \ref{fig:PulseSequence} top), separately integrated over the window of highest amplifier SNR (see Figure \ref{fig:PulseSequence} bottom) to get a single value for signal and reference then divided for contrast .

%% file: References.bib
@article{nishikawa_coherent_2025,
	title = {Coherent photoelectrical readout of single spins in silicon carbide at room temperature},
	volume = {16},
	issn = {2041-1723},
	url = {https://www.nature.com/articles/s41467-025-58629-1},
	doi = {10.1038/s41467-025-58629-1},
	language = {en},
	number = {1},
	urldate = {2025-05-07},
	journal = {Nat. Commun.},
	author = {Nishikawa, T. and Morioka, N. and Abe, H. and Murata, K. and Okajima, K. and Ohshima, T. and Tsuchida, H. and Mizuochi, N.},
	month = apr,
	year = {2025},
	pages = {3405},
}

@article{hu_room-temperature_2024,
	title = {Room-temperature waveguide integrated quantum register in a semiconductor photonic platform},
	volume = {15},
	issn = {2041-1723},
	url = {https://www.nature.com/articles/s41467-024-54606-2},
	doi = {10.1038/s41467-024-54606-2},
	language = {en},
	number = {1},
	urldate = {2025-03-18},
	journal = {Nat. Commun.},
	author = {Hu, H. and Zhou, Y. and Yi, A. and Bao, T. and Liu, C. and Luo, Q. and Zhang, Y. and Wang, Z. and Li, Q. and Lu, D. and Liu, Z. and Xiao, S. and Ou, X. and Song, Q.},
	month = nov,
	year = {2024},
	pages = {10256},
}

@article{nishikawa_electrical_2022,
	title = {Electrical detection of nuclear spins via silicon vacancies in silicon carbide at room temperature},
	volume = {121},
	issn = {0003-6951, 1077-3118},
	url = {https://pubs.aip.org/apl/article/121/18/184005/2834789/Electrical-detection-of-nuclear-spins-via-silicon},
	doi = {10.1063/5.0115928},
	language = {en},
	number = {18},
	urldate = {2024-08-07},
	journal = {Appl. Phys. Lett.},
	author = {Nishikawa, T. and Morioka, N. and Abe, H. and Morishita, H. and Ohshima, T. and Mizuochi, N.},
	month = oct,
	year = {2022},
	pages = {184005},
}

@article{lipton_low-loss_2025,
	title = {Low-{Loss} {Nanophotonic} {Devices} with {Chip}-{Level} {Uniformity} and {Integrated} {Color} {Centers} in {SiC}-{On}-{Insulator}},
	volume = {12},
	copyright = {https://doi.org/10.15223/policy-029},
	issn = {2330-4022, 2330-4022},
	url = {https://pubs.acs.org/doi/10.1021/acsphotonics.4c01834},
	doi = {10.1021/acsphotonics.4c01834},
	language = {en},
	number = {5},
	urldate = {2025-11-05},
	journal = {ACS Photonics},
	author = {Lipton, J. and Yurash, B. and Sorensen, A. and Vajo, J. and Whiteley, S. and Wang, T. and Huang, B. and Bai, X. and Portales, A. and Rubin, S. and Strayer, J. and Graetz, J. and Cui, S.},
	month = may,
	year = {2025},
	pages = {2397--2405},
}

@article{wang_high-q_2021,
	title = {High-{Q} microresonators on {4H}-silicon-carbide-on-insulator platform for nonlinear photonics},
	volume = {10},
	issn = {2047-7538},
	url = {https://www.nature.com/articles/s41377-021-00584-9},
	doi = {10.1038/s41377-021-00584-9},
	language = {en},
	number = {1},
	urldate = {2025-11-05},
	journal = {Light-Sci. Appl.},
	author = {Wang, C. and Fang, Z. and Yi, A. and Yang, B. and Wang, Z. and Zhou, L. and Shen, C. and Zhu, Y. and Zhou, Y. and Bao, R. and Li, Z. and Chen, Y. and Huang, K. and Zhang, J. and Cheng, Y. and Ou, X.},
	month = jul,
	year = {2021},
	pages = {139},
}

@article{wolfowicz_optical_2017,
	title = {Optical charge state control of spin defects in {4H}-{SiC}},
	volume = {8},
	issn = {2041-1723},
	url = {https://www.nature.com/articles/s41467-017-01993-4},
	doi = {10.1038/s41467-017-01993-4},
	language = {en},
	number = {1},
	urldate = {2025-11-05},
	journal = {Nat. Commun.},
	author = {Wolfowicz, G. and Anderson, C. P. and Yeats, A. L. and Whiteley, S. J. and Niklas, J. and Poluektov, O. G. and Heremans, F. J. and Awschalom, D. D.},
	month = nov,
	year = {2017},
	pages = {1876},
}

@article{son_charge_2021,
	title = {Charge state control of the silicon vacancy and divacancy in silicon carbide},
	volume = {129},
	issn = {0021-8979, 1089-7550},
	url = {https://pubs.aip.org/jap/article/129/21/215702/1075273/Charge-state-control-of-the-silicon-vacancy-and},
	doi = {10.1063/5.0052131},
	language = {en},
	number = {21},
	urldate = {2025-11-05},
	journal = {J. Appl. Phys.},
	author = {Son, N. T. and Ivanov, I. G.},
	month = jun,
	year = {2021},
	pages = {215702},
}

@article{baranov_silicon_2011,
	title = {Silicon vacancy in {SiC} as a promising quantum system for single-defect and single-photon spectroscopy},
	volume = {83},
	copyright = {http://link.aps.org/licenses/aps-default-license},
	issn = {1098-0121, 1550-235X},
	url = {https://link.aps.org/doi/10.1103/PhysRevB.83.125203},
	doi = {10.1103/PhysRevB.83.125203},
	language = {en},
	number = {12},
	urldate = {2025-11-05},
	journal = {Phys. Rev. B},
	author = {Baranov, P. G. and Bundakova, A. P. and Soltamova, A. A. and Orlinskii, S. B. and Borovykh, I. V. and Zondervan, Rob and Verberk, R. and Schmidt, J.},
	month = mar,
	year = {2011},
	pages = {125203},
}

@article{shang_local_2020,
	title = {Local vibrational modes of {Si} vacancy spin qubits in {SiC}},
	volume = {101},
	issn = {2469-9950, 2469-9969},
	url = {https://link.aps.org/doi/10.1103/PhysRevB.101.144109},
	doi = {10.1103/PhysRevB.101.144109},
	language = {en},
	number = {14},
	urldate = {2025-11-05},
	journal = {Phys. Rev. B},
	author = {Shang, Z. and Hashemi, A. and Berencén, Y. and Komsa, H.-P. and Erhart, P. and Zhou, S. and Helm, M. and Krasheninnikov, A. V. and Astakhov, G. V.},
	month = apr,
	year = {2020},
	pages = {144109},
}

@article{fuchs_engineering_2015,
	title = {Engineering near-infrared single-photon emitters with optically active spins in ultrapure silicon carbide},
	volume = {6},
	issn = {2041-1723},
	url = {https://www.nature.com/articles/ncomms8578},
	doi = {10.1038/ncomms8578},
	language = {en},
	number = {1},
	urldate = {2025-11-05},
	journal = {Nat. Commun.},
	author = {Fuchs, F. and Stender, B. and Trupke, M. and Simin, D. and Pflaum, J. and Dyakonov, V. and Astakhov, G. V.},
	month = jul,
	year = {2015},
	pages = {7578},
}

@article{udvarhelyi_vibronic_2020,
	title = {Vibronic {States} and {Their} {Effect} on the {Temperature} and {Strain} {Dependence} of {Silicon}-{Vacancy} {Qubits} in {4H}-{SiC}},
	volume = {13},
	issn = {2331-7019},
	url = {https://link.aps.org/doi/10.1103/PhysRevApplied.13.054017},
	doi = {10.1103/PhysRevApplied.13.054017},
	language = {en},
	number = {5},
	urldate = {2025-11-05},
	journal = {Phys. Rev. Applied},
	author = {Udvarhelyi, P. and Thiering, G. and Morioka, N. and Babin, C. and Kaiser, F. and Lukin, D. and Ohshima, T. and Ul-Hassan, J. and Son, N. T. and Vučković, J. and Wrachtrup, J. and Gali, A.},
	month = may,
	year = {2020},
	pages = {054017},
}

@article{li_non-invasive_2025,
	title = {Non-invasive bioinert room-temperature quantum sensor from silicon carbide qubits},
	issn = {1476-1122, 1476-4660},
	url = {https://www.nature.com/articles/s41563-025-02382-9},
	doi = {10.1038/s41563-025-02382-9},
	language = {en},
	urldate = {2025-11-06},
	journal = {Nat. Mater.},
	author = {Li, P. and Zhou, J. and Li, S. and Udvarhelyi, P. and Xu, J. and Li, C. and Huang, B. and Guo, G. and Gali, A.},
	month = nov,
	year = {2025},
}

@article{powell_integrated_2022,
	title = {Integrated silicon carbide electro-optic modulator},
	volume = {13},
	issn = {2041-1723},
	url = {https://www.nature.com/articles/s41467-022-29448-5},
	doi = {10.1038/s41467-022-29448-5},
	language = {en},
	number = {1},
	urldate = {2025-11-06},
	journal = {Nat. Commun.},
	author = {Powell, K. and Li, L. and Shams-Ansari, A. and Wang, J. and Meng, D. and Sinclair, N. and Deng, J. and Lončar, M. and Yi, X.},
	month = apr,
	year = {2022},
	pages = {1851},
}

@article{christle_isolated_2015,
	title = {Isolated electron spins in silicon carbide with millisecond coherence times},
	volume = {14},
	issn = {1476-1122, 1476-4660},
	url = {https://www.nature.com/articles/nmat4144},
	doi = {10.1038/nmat4144},
	language = {en},
	number = {2},
	urldate = {2025-11-14},
	journal = {Nat. Mater.},
	author = {Christle, D. J. and Falk, A. L. and Andrich, P. and Klimov, P. V. and Hassan, J. Ul. and Son, N. T. and Janzén, E. and Ohshima, T. and Awschalom, D. D.},
	month = feb,
	year = {2015},
	pages = {160--163},
}

@article{baranov_epr_2005,
	title = {{EPR} identification of the triplet ground state and photoinduced population inversion for a {Si}-{C} divacancy in silicon carbide},
	volume = {82},
	copyright = {http://www.springer.com/tdm},
	issn = {0021-3640, 1090-6487},
	url = {http://link.springer.com/10.1134/1.2142873},
	doi = {10.1134/1.2142873},
	language = {en},
	number = {7},
	urldate = {2025-11-14},
	journal = {JEPT Letters.},
	author = {Baranov, P. G. and Il’in, I. V. and Mokhov, E. N. and Muzafarova, M. V. and Orlinskii, S. B. and Schmidt, J.},
	month = oct,
	year = {2005},
	pages = {441--443},
}

@article{widmann_coherent_2015,
	title = {Coherent control of single spins in silicon carbide at room temperature},
	volume = {14},
	issn = {1476-1122, 1476-4660},
	url = {https://www.nature.com/articles/nmat4145},
	doi = {10.1038/nmat4145},
	language = {en},
	number = {2},
	urldate = {2025-11-14},
	journal = {Nat. Mater.},
	author = {Widmann, M. and Lee, S. and Rendler, T. and Son, N. T. and Fedder, H. and Paik, S. and Yang, L. and Zhao, N. and Yang, S. and Booker, I. and Denisenko, A. and Jamali, M. and Momenzadeh, S. A. and Gerhardt, I. and Ohshima, T. and Gali, A. and Janzén, E. and Wrachtrup, J.},
	month = feb,
	year = {2015},
	pages = {164--168},
}

@article{radulaski_scalable_2017,
	title = {Scalable {Quantum} {Photonics} with {Single} {Color} {Centers} in {Silicon} {Carbide}},
	volume = {17},
	issn = {1530-6984, 1530-6992},
	url = {https://pubs.acs.org/doi/10.1021/acs.nanolett.6b05102},
	doi = {10.1021/acs.nanolett.6b05102},
	language = {en},
	number = {3},
	urldate = {2025-11-14},
	journal = {Nano Lett.},
	author = {Radulaski, M. and Widmann, M. and Niethammer, M. and Zhang, J. L. and Lee, S. and Rendler, T. and Lagoudakis, K. G. and Son, N. T. and Janzén, E. and Ohshima, T. and Wrachtrup, J. and Vučković, J.},
	month = mar,
	year = {2017},
	pages = {1782--1786},
}

@article{mizuochi_continuous-wave_2002,
	title = {Continuous-wave and pulsed {EPR} study of the negatively charged silicon vacancy with {S} = 3/2 and {C3v} symmetry in n-type {4H}-{SiC}},
	volume = {66},
	copyright = {http://link.aps.org/licenses/aps-default-license},
	issn = {0163-1829, 1095-3795},
	url = {https://link.aps.org/doi/10.1103/PhysRevB.66.235202},
	doi = {10.1103/PhysRevB.66.235202},
	language = {en},
	number = {23},
	urldate = {2025-11-14},
	journal = {Phys. Rev. B},
	author = {Mizuochi, N. and Yamasaki, S. and Takizawa, H. and Morishita, N. and Ohshima, T. and Itoh, H. and Isoya, J.},
	month = dec,
	year = {2002},
	pages = {235202},
}

@article{wang_experimental_2020,
	title = {Experimental {Optical} {Properties} of {Single} {Nitrogen} {Vacancy} {Centers} in {Silicon} {Carbide} at {Room} {Temperature}},
	volume = {7},
	copyright = {https://doi.org/10.15223/policy-029},
	issn = {2330-4022, 2330-4022},
	url = {https://pubs.acs.org/doi/10.1021/acsphotonics.0c00218},
	doi = {10.1021/acsphotonics.0c00218},
	language = {en},
	number = {7},
	urldate = {2025-11-14},
	journal = {ACS Photonics},
	author = {Wang, J. and Liu, Z. and Yan, F. and Li, Q. and Yang, X. and Guo, L. and Zhou, X. and Huang, W. and Xu, J. and Li, C. and Guo, G.},
	month = jul,
	year = {2020},
	pages = {1611--1616},
}

@article{norman_sub-2_2025,
	title = {Sub-2 {Kelvin} {Characterization} of {Nitrogen}-{Vacancy} {Centers} in {Silicon} {Carbide} {Nanopillars}},
	volume = {12},
	copyright = {https://creativecommons.org/licenses/by-nc-nd/4.0/},
	issn = {2330-4022, 2330-4022},
	url = {https://pubs.acs.org/doi/10.1021/acsphotonics.5c00096},
	doi = {10.1021/acsphotonics.5c00096},
	language = {en},
	number = {5},
	urldate = {2025-11-14},
	journal = {ACS Photonics},
	author = {Norman, V. A. and Majety, S. and Rubin, A. H. and Saha, P. and Gonzalez, N. R. and Simo, J. and Palomarez, . and Li, L. and Curro, P. B. and Dhuey, S. and Virasawmy, S. and Radulaski, M.},
	month = may,
	year = {2025},
	pages = {2604--2611},
}

@article{wang_coherent_2020,
	title = {Coherent {Control} of {Nitrogen}-{Vacancy} {Center} {Spins} in {Silicon} {Carbide} at {Room} {Temperature}},
	volume = {124},
	issn = {0031-9007, 1079-7114},
	url = {https://link.aps.org/doi/10.1103/PhysRevLett.124.223601},
	doi = {10.1103/PhysRevLett.124.223601},
	language = {en},
	number = {22},
	urldate = {2025-11-14},
	journal = {Phys. Rev. Lett.},
	author = {Wang, J. and Yan, F. and Li, Q. and Liu, Z. and Liu, H. and Guo, G. and Guo, L. and Zhou, X. and Cui, J. and Wang, J. and Zhou, Z. and Xu, X. and Xu, J. and Li, C. and Guo, G.},
	month = jun,
	year = {2020},
	pages = {223601},
}

@article{simin_all-optical_2016,
	title = {All-{Optical} dc {Nanotesla} {Magnetometry} {Using} {Silicon} {Vacancy} {Fine} {Structure} in {Isotopically} {Purified} {Silicon} {Carbide}},
	volume = {6},
	copyright = {http://creativecommons.org/licenses/by/3.0/},
	issn = {2160-3308},
	url = {https://link.aps.org/doi/10.1103/PhysRevX.6.031014},
	doi = {10.1103/PhysRevX.6.031014},
	language = {en},
	number = {3},
	urldate = {2025-11-14},
	journal = {Phys. Rev. X},
	author = {Simin, D. and Soltamov, V. A. and Poshakinskiy, A. V. and Anisimov, A. N. and Babunts, R. A. and Tolmachev, D. O. and Mokhov, E. N. and Trupke, M. and Tarasenko, S. A. and Sperlich, A. and Baranov, P. G. and Dyakonov, V. and Astakhov, G. V.},
	month = jul,
	year = {2016},
	pages = {031014},
}

@article{tahara_quantum_2025,
	title = {Quantum sensing with duplex qubits of silicon vacancy centers in {SiC} at room temperature},
	volume = {11},
	issn = {2056-6387},
	url = {https://www.nature.com/articles/s41534-025-01011-2},
	doi = {10.1038/s41534-025-01011-2},
	language = {en},
	number = {1},
	urldate = {2025-11-14},
	journal = {npj Quantum Inf.},
	author = {Tahara, K. and Tamura, S. and Toyama, H. and Nakane, J. J. and Kutsuki, K. and Yamazaki, Y. and Ohshima, T.},
	month = apr,
	year = {2025},
	pages = {58},
}

@article{castelletto_quantum_2024,
	title = {Quantum systems in silicon carbide for sensing applications},
	volume = {87},
	issn = {0034-4885, 1361-6633},
	url = {https://iopscience.iop.org/article/10.1088/1361-6633/ad10b3},
	doi = {10.1088/1361-6633/ad10b3},
	language = {en},
	number = {1},
	urldate = {2025-11-14},
	journal = {Rep. Prog. Phys.},
	author = {Castelletto, S. and Lew, C. T-K. and Lin, W. and Xu, J.},
	month = jan,
	year = {2024},
	pages = {014501},
}

@article{nagy_high-fidelity_2019,
	title = {High-fidelity spin and optical control of single silicon-vacancy centres in silicon carbide},
	volume = {10},
	issn = {2041-1723},
	url = {https://www.nature.com/articles/s41467-019-09873-9},
	doi = {10.1038/s41467-019-09873-9},
	language = {en},
	number = {1},
	urldate = {2025-11-14},
	journal = {Nat. Commun.},
	author = {Nagy, R. and Niethammer, M. and Widmann, M. and Chen, Y. and Udvarhelyi, P. and Bonato, C. and Hassan, J. U. and Karhu, R. and Ivanov, I. G. and Son, N. T. and Maze, J. R. and Ohshima, T. and Soykal, Ö. O. and Gali, Á. and Lee, S. and Kaiser, F. and Wrachtrup, J.},
	month = apr,
	year = {2019},
	pages = {1954},
}

@article{castelletto_silicon_2020,
	title = {Silicon carbide color centers for quantum applications},
	volume = {2},
	issn = {2515-7647},
	url = {https://iopscience.iop.org/article/10.1088/2515-7647/ab77a2},
	doi = {10.1088/2515-7647/ab77a2},
	language = {en},
	number = {2},
	urldate = {2025-11-14},
	journal = {J. Phys. Photonics},
	author = {Castelletto, S. and Boretti, A.},
	month = apr,
	year = {2020},
	pages = {022001},
}

@article{xu_silicon_2021,
	title = {Silicon carbide based quantum networking},
	volume = {1},
	issn = {26673258},
	url = {https://linkinghub.elsevier.com/retrieve/pii/S2667325820300042},
	doi = {10.1016/j.fmre.2020.11.004},
	language = {en},
	number = {2},
	urldate = {2025-11-14},
	journal = {Fundamental Research},
	author = {Xu, J. and Li, C. and Guo, G.},
	month = mar,
	year = {2021},
	pages = {220--222},
}

@article{bader_analysis_2024,
	title = {Analysis, recent challenges and capabilities of spin-photon interfaces in {Silicon} carbide-on-insulator},
	volume = {1},
	issn = {2948-216X},
	url = {https://www.nature.com/articles/s44310-024-00031-8},
	doi = {10.1038/s44310-024-00031-8},
	language = {en},
	number = {1},
	urldate = {2025-11-14},
	journal = {npj Nanophoton.},
	author = {Bader, J. and Arianfard, H. and Peruzzo, A. and Castelletto, S.},
	month = aug,
	year = {2024},
	pages = {29},
}

@article{song_ultrahigh-q_2019,
	title = {Ultrahigh-{Q} photonic crystal nanocavities based on {4H} silicon carbide},
	volume = {6},
	issn = {2334-2536},
	url = {https://opg.optica.org/abstract.cfm?URI=optica-6-8-991},
	doi = {10.1364/OPTICA.6.000991},
	language = {en},
	number = {8},
	urldate = {2025-11-14},
	journal = {Optica},
	author = {Song, B. and Asano, T. and Jeon, S. and Kim, H. and Chen, C. and Kang, D. D. and Noda, S.},
	month = aug,
	year = {2019},
	pages = {991},
}

@article{burkard_spin-entangled_2007,
	title = {Spin-entangled electrons in solid-state systems},
	volume = {19},
	issn = {0953-8984, 1361-648X},
	url = {https://iopscience.iop.org/article/10.1088/0953-8984/19/23/233202},
	doi = {10.1088/0953-8984/19/23/233202},
	language = {en},
	number = {23},
	urldate = {2025-11-14},
	journal = {J. Phys.: Condens. Matter.},
	author = {Burkard, G.},
	month = jun,
	year = {2007},
	pages = {233202},
}

@article{main_distributed_2025,
	title = {Distributed quantum computing across an optical network link},
	volume = {638},
	issn = {0028-0836, 1476-4687},
	url = {https://www.nature.com/articles/s41586-024-08404-x},
	doi = {10.1038/s41586-024-08404-x},
	language = {en},
	number = {8050},
	urldate = {2025-11-14},
	journal = {Nature},
	author = {Main, D. and Drmota, P. and Nadlinger, D. P. and Ainley, E. M. and Agrawal, A. and Nichol, B. C. and Srinivas, R. and Araneda, G. and Lucas, D. M.},
	month = feb,
	year = {2025},
	pages = {383--388},
}

@article{kraus_magnetic_2014,
	title = {Magnetic field and temperature sensing with atomic-scale spin defects in silicon carbide},
	volume = {4},
	issn = {2045-2322},
	url = {https://www.nature.com/articles/srep05303},
	doi = {10.1038/srep05303},
	language = {en},
	number = {1},
	urldate = {2025-11-15},
	journal = {Sci. Rep.},
	author = {Kraus, H. and Soltamov, V. A. and Fuchs, F. and Simin, D. and Sperlich, A. and Baranov, P. G. and Astakhov, G. V. and Dyakonov, V.},
	month = jul,
	year = {2014},
	pages = {5303},
}

@article{siyushev_photoelectrical_2019,
	title = {Photoelectrical imaging and coherent spin-state readout of single nitrogen-vacancy centers in diamond},
	volume = {363},
	issn = {0036-8075, 1095-9203},
	url = {https://www.science.org/doi/10.1126/science.aav2789},
	doi = {10.1126/science.aav2789},
	language = {en},
	number = {6428},
	urldate = {2025-11-17},
	journal = {Science},
	author = {Siyushev, P. and Nesladek, M. and Bourgeois, E. and Gulka, M. and Hruby, J. and Yamamoto, T. and Trupke, M. and Teraji, T. and Isoya, J. and Jelezko, F.},
	month = feb,
	year = {2019},
	pages = {728--731},
}

@article{bourgeois_photoelectric_2015,
	title = {Photoelectric detection of electron spin resonance of nitrogen-vacancy centres in diamond},
	volume = {6},
	issn = {2041-1723},
	url = {https://www.nature.com/articles/ncomms9577},
	doi = {10.1038/ncomms9577},
	language = {en},
	number = {1},
	urldate = {2025-11-17},
	journal = {Nat. Commun.},
	author = {Bourgeois, E. and Jarmola, A. and Siyushev, P. and Gulka, M. and Hruby, J. and Jelezko, F. and Budker, D. and Nesladek, M.},
	month = oct,
	year = {2015},
	pages = {8577},
}

@article{ru_room-temperature_2025,
	title = {Room-{Temperature} {Electrical} {Readout} of {Spin} {Defects} in van der {Waals} {Materials}},
	url = {https://journals.aps.org/prl/accepted/10.1103/dlzw-dhsr},
	doi = {10.1103/dlzw-dhsr},
	language = {en},
	urldate = {2025-11-17},
	journal = {Phys. Rev. Lett.},
	author = {Ru, S. and An, L. and Liang, H. and Jiang, Z. and Li, Z. and Lyu, X. and Zhou, F. and Cai, H. and Yang, Y. and He, R. and Cernansky, R. and Teo, E. H. T. and Mukherjee, M. and Bettiol, A. A. and Zúñiga-Perez, J. and Jelezko, F. and Gao, W.},
	month = nov,
	year = {2025},
}

@article{lew_all-electrical_2024-1,
	title = {All-{Electrical} {Readout} of {Coherently} {Controlled} {Spins} in {Silicon} {Carbide}},
	volume = {132},
	issn = {0031-9007, 1079-7114},
	url = {https://link.aps.org/doi/10.1103/PhysRevLett.132.146902},
	doi = {10.1103/PhysRevLett.132.146902},
	language = {en},
	number = {14},
	urldate = {2025-11-17},
	journal = {Phys. Rev. Lett.},
	author = {Lew, C. T.-K. and Sewani, V. K. and Iwamoto, N. and Ohshima, T. and McCallum, J. C. and Johnson, B. C.},
	month = apr,
	year = {2024},
	pages = {146902},
}

@article{koppens_driven_2006,
	title = {Driven coherent oscillations of a single electron spin in a quantum dot},
	volume = {442},
	copyright = {http://www.springer.com/tdm},
	issn = {0028-0836, 1476-4687},
	url = {https://www.nature.com/articles/nature05065},
	doi = {10.1038/nature05065},
	language = {en},
	number = {7104},
	urldate = {2025-11-17},
	journal = {Nature},
	author = {Koppens, F. H. L. and Buizert, C. and Tielrooij, K. J. and Vink, I. T. and Nowack, K. C. and Meunier, T. and Kouwenhoven, L. P. and Vandersypen, L. M. K.},
	month = aug,
	year = {2006},
	pages = {766--771},
}

@article{briegel_optical_2025,
	title = {Optical widefield nuclear magnetic resonance microscopy},
	volume = {16},
	issn = {2041-1723},
	url = {https://www.nature.com/articles/s41467-024-55003-5},
	doi = {10.1038/s41467-024-55003-5},
	language = {en},
	number = {1},
	urldate = {2025-11-18},
	journal = {Nat. Commun.},
	author = {Briegel, K. D. and Von Grafenstein, N. R. and Draeger, J. C. and Blümler, P. and Allert, R. D. and Bucher, D. B.},
	month = feb,
	year = {2025},
	pages = {1281},
}

@article{fisher_high-resolution_2025,
	title = {High-{Resolution} {Nanoscale} {AC} {Quantum} {Sensing} in {CMOS} {Compatible} {SiC}},
	volume = {25},
	copyright = {https://creativecommons.org/licenses/by/4.0/},
	issn = {1530-6984, 1530-6992},
	url = {https://pubs.acs.org/doi/10.1021/acs.nanolett.5c02515},
	doi = {10.1021/acs.nanolett.5c02515},
	language = {en},
	number = {30},
	urldate = {2025-11-19},
	journal = {Nano Lett.},
	author = {Fisher, P. and Zappacosta, A. and Fuhrmann, J. and Haylock, B. and Gao, W. and Nagy, R. and Jelezko, F. and Cernansky, R.},
	month = jul,
	year = {2025},
	pages = {11626--11631},
}

@article{niethammer_coherent_2019,
	title = {Coherent electrical readout of defect spins in silicon carbide by photo-ionization at ambient conditions},
	volume = {10},
	issn = {2041-1723},
	url = {https://www.nature.com/articles/s41467-019-13545-z},
	doi = {10.1038/s41467-019-13545-z},
	language = {en},
	number = {1},
	urldate = {2025-11-19},
	journal = {Nat. Commun.},
	author = {Niethammer, M. and Widmann, M. and Rendler, T. and Morioka, N. and Chen, Y. and Stöhr, R. and Hassan, J. U. and Onoda, S. and Ohshima, T. and Lee, S. and Mukherjee, A. and Isoya, J. and Son, N. T. and Wrachtrup, J.},
	month = dec,
	year = {2019},
	pages = {5569},
}

@article{stegner_electrical_2006,
	title = {Electrical detection of coherent {31P} spin quantum states},
	volume = {2},
	issn = {1745-2473, 1745-2481},
	url = {https://www.nature.com/articles/nphys465},
	doi = {10.1038/nphys465},
	language = {en},
	number = {12},
	urldate = {2025-11-22},
	journal = {Nature Phys},
	author = {Stegner, Andre R. and Boehme, Christoph and Huebl, Hans and Stutzmann, Martin and Lips, Klaus and Brandt, Martin S.},
	month = dec,
	year = {2006},
	pages = {835--838},
}

@article{abe_electrical_2022,
	title = {Electrical detection of \textit{{T}} {V2a}-type silicon vacancy spin defect in {4H}-{SiC} {MOSFETs}},
	volume = {120},
	issn = {0003-6951, 1077-3118},
	url = {https://pubs.aip.org/apl/article/120/6/064001/2833306/Electrical-detection-of-TV2a-type-silicon-vacancy},
	doi = {10.1063/5.0078189},
	language = {en},
	number = {6},
	urldate = {2025-11-25},
	journal = {Applied Physics Letters},
	author = {Abe, Yuta and Chaen, Akihumi and Sometani, Mitsuru and Harada, Shinsuke and Yamazaki, Yuichi and Ohshima, Takeshi and Umeda, Takahide},
	month = feb,
	year = {2022},
	pages = {064001},
}

@article{carter_spin_2015,
	title = {Spin coherence and echo modulation of the silicon vacancy in 4 {H} − {SiC} at room temperature},
	volume = {92},
	copyright = {http://link.aps.org/licenses/aps-default-license},
	issn = {1098-0121, 1550-235X},
	url = {https://link.aps.org/doi/10.1103/PhysRevB.92.161202},
	doi = {10.1103/PhysRevB.92.161202},
	language = {en},
	number = {16},
	urldate = {2025-11-26},
	journal = {Phys. Rev. B},
	author = {Carter, S. G. and Soykal, Ö. O. and Dev, Pratibha and Economou, Sophia E. and Glaser, E. R.},
	month = oct,
	year = {2015},
	pages = {161202},
}

@article{hrubesch_efficient_2017,
	title = {Efficient {Electrical} {Spin} {Readout} of {NV} − {Centers} in {Diamond}},
	volume = {118},
	copyright = {http://link.aps.org/licenses/aps-default-license},
	issn = {0031-9007, 1079-7114},
	url = {https://link.aps.org/doi/10.1103/PhysRevLett.118.037601},
	doi = {10.1103/PhysRevLett.118.037601},
	language = {en},
	number = {3},
	urldate = {2025-11-26},
	journal = {Phys. Rev. Lett.},
	author = {Hrubesch, Florian M. and Braunbeck, Georg and Stutzmann, Martin and Reinhard, Friedemann and Brandt, Martin S.},
	month = jan,
	year = {2017},
	pages = {037601},
}

@article{steidl_single_2025,
	title = {Single {V2} defect in {4H} silicon carbide {Schottky} diode at low temperature},
	volume = {16},
	issn = {2041-1723},
	url = {https://www.nature.com/articles/s41467-025-59647-9},
	doi = {10.1038/s41467-025-59647-9},
	language = {en},
	number = {1},
	urldate = {2025-11-26},
	journal = {Nat Commun},
	author = {Steidl, Timo and Kuna, Pierre and Hesselmeier-Hüttmann, Erik and Liu, Di and Stöhr, Rainer and Knolle, Wolfgang and Ghezellou, Misagh and Ul-Hassan, Jawad and Schober, Maximilian and Bockstedte, Michel and Bian, Guodong and Gali, Adam and Vorobyov, Vadim and Wrachtrup, Jörg},
	month = may,
	year = {2025},
	pages = {4669},
}

@article{bulancea-lindvall_temperature_2024,
	title = {Temperature dependence of the {AB} lines and optical properties of the carbon–antisite-vacancy pair in 4 {H} − {Si} {C}},
	volume = {22},
	issn = {2331-7019},
	url = {https://link.aps.org/doi/10.1103/PhysRevApplied.22.034056},
	doi = {10.1103/PhysRevApplied.22.034056},
	language = {en},
	number = {3},
	urldate = {2026-02-18},
	journal = {Phys. Rev. Applied},
	author = {Bulancea-Lindvall, Oscar and Davidsson, Joel and Ivanov, Ivan G. and Gali, Adam and Ivády, Viktor and Armiento, Rickard and Abrikosov, Igor A.},
	month = sep,
	year = {2024},
	pages = {034056},
}

@article{nakane_deep_2021,
	title = {Deep levels related to the carbon antisite–vacancy pair in {4H}-{SiC}},
	volume = {130},
	issn = {0021-8979, 1089-7550},
	url = {https://pubs.aip.org/jap/article/130/6/065703/346764/Deep-levels-related-to-the-carbon-antisite-vacancy},
	doi = {10.1063/5.0059953},
	language = {en},
	number = {6},
	urldate = {2026-02-18},
	journal = {Journal of Applied Physics},
	author = {Nakane, Hiroki and Kato, Masashi and Ohkouchi, Yutaro and Trinh, Xuan Thang and Ivanov, Ivan G. and Ohshima, Takeshi and Son, Nguyen Tien},
	month = aug,
	year = {2021},
	pages = {065703},
}

@article{wang_robust_2021,
	title = {Robust coherent control of solid-state spin qubits using anti-{Stokes} excitation},
	volume = {12},
	issn = {2041-1723},
	url = {https://www.nature.com/articles/s41467-021-23471-8},
	doi = {10.1038/s41467-021-23471-8},
	language = {en},
	number = {1},
	urldate = {2026-02-17},
	journal = {Nat Commun},
	author = {Wang, Jun-Feng and Yan, Fei-Fei and Li, Qiang and Liu, Zheng-Hao and Cui, Jin-Ming and Liu, Zhao-Di and Gali, Adam and Xu, Jin-Shi and Li, Chuan-Feng and Guo, Guang-Can},
	month = may,
	year = {2021},
	pages = {3223},
}

@article{nagy_quantum_2018,
	title = {Quantum {Properties} of {Dichroic} {Silicon} {Vacancies} in {Silicon} {Carbide}},
	volume = {9},
	issn = {2331-7019},
	url = {https://link.aps.org/doi/10.1103/PhysRevApplied.9.034022},
	doi = {10.1103/PhysRevApplied.9.034022},
	language = {en},
	number = {3},
	urldate = {2026-02-17},
	journal = {Phys. Rev. Applied},
	author = {Nagy, Roland and Widmann, Matthias and Niethammer, Matthias and Dasari, Durga B. R. and Gerhardt, Ilja and Soykal, Öney O. and Radulaski, Marina and Ohshima, Takeshi and Vučković, Jelena and Son, Nguyen Tien and Ivanov, Ivan G. and Economou, Sophia E. and Bonato, Cristian and Lee, Sang-Yun and Wrachtrup, Jörg},
	month = mar,
	year = {2018},
	pages = {034022},
}

@article{okajima_photoionization_2026,
	title = {Photoionization {Current} {Spectroscopy} of {Individual} {Silicon} {Vacancies} in {Silicon} {Carbide}},
	copyright = {https://doi.org/10.15223/policy-029},
	issn = {1530-6984, 1530-6992},
	url = {https://pubs.acs.org/doi/10.1021/acs.nanolett.5c05964},
	doi = {10.1021/acs.nanolett.5c05964},
	language = {en},
	urldate = {2026-04-21},
	journal = {Nano Lett.},
	author = {Okajima, Kazuki and Nishikawa, Tetsuri and Abe, Hiroshi and Murata, Koichi and Ohshima, Takeshi and Tsuchida, Hidekazu and Morioka, Naoya and Mizuochi, Norikazu},
	month = apr,
	year = {2026},
	pages = {acs.nanolett.5c05964},
}
